\documentclass[aps,pre,reprint,groupedaddress,nofootinbib]{revtex4-1}

\usepackage{amsmath}
\usepackage{amssymb}
\usepackage[figuresright]{rotating}
\usepackage{color}
\usepackage{graphicx}
\usepackage[justification=justified,singlelinecheck=false]{caption}
\usepackage{subcaption}
\usepackage[sort&compress]{natbib}
\usepackage{changes} %for track changes

\renewcommand{\v}[1]{\ensuremath{\mathbf{#1}}} % for vectors
\newcommand{\gv}[1]{\ensuremath{\mbox{\boldmath$ #1 $}}}
\newcommand{\ket}[1]{\left| #1 \right>} % for Dirac bras
\newcommand{\be}{\begin{equation}}
\newcommand{\ee}{\end{equation}}
\newcommand{\bea}{\begin{eqnarray}}
\newcommand{\eea}{\end{eqnarray}}

\def\lab{\label}

\def\pa{\partial}

\def\al{\alpha}

\def\ga{\gamma}

\def\de{\delta}

\def\ep{\epsilon}

\def\la{\lambda}

\def\si{\sigma}

\def\om{\omega}
\def\Om{\Omega}

\begin{document}

\title{Water-mediated correlations in DNA-enzyme interactions
%Or: Dipole wave-mediated correlations in DNA-enzyme interactions
} %Title of paper

\author{P. Kurian}
\email[Corresponding author. E-mail: ]{pkurian@howard.edu.}
%\email[]{}
%\homepage[]{Your web page}
%\thanks{}
\affiliation{National Human Genome Center and Department of Medicine, Howard University College of Medicine, Washington, DC 20059, USA; Computational Physics Laboratory, Howard University, Washington, DC 20059, USA}

\author{A. Capolupo}
%\homepage[]{Your web page}
%\thanks{}
\affiliation{Universit\`{a} degli Studi di Salerno and INFN Gruppo Collegato di Salerno, 84084 Fisciano (Salerno), Italy}

\author{T. J. A. Craddock}
\affiliation{Departments of Psychology and Neuroscience, Computer Science, and Clinical Immunology, and Clinical Systems Biology Group, Institute for Neuro-Immune Medicine, Nova Southeastern University, Fort Lauderdale, FL 33314, USA}

\author{G. Vitiello}
%\email[]{}
%\homepage[]{Your web page}
%\thanks{}
\affiliation{Universit\`{a} degli Studi di Salerno and INFN Gruppo Collegato di Salerno, 84084 Fisciano (Salerno), Italy }

\date{\today}

\begin{abstract}
In this paper we consider dipole-mediated correlations between DNA and enzymes in the context of their water environment. Such correlations emerge from electric dipole-dipole interactions between aromatic ring structures in DNA and in enzymes. We show that there are matching collective modes between DNA and enzyme dipole fields, and that a dynamic time-averaged polarization vanishes in the water dipole field only if either DNA, enzyme, or both are absent from the sample. This persistent field may serve as the electromagnetic image that, in popular colloquialisms about DNA biochemistry, allows enzymes to ``scan'' or ``read'' the double helix. Topologically nontrivial configurations in the coherent ground state requiring clamplike enzyme behavior on the DNA may stem, ultimately, from spontaneously broken gauge symmetries.
\end{abstract}

%\pacs{11.10.---z, 12.20.---m, 64.90.+b, 11.30.Qc}% insert suggested PACS numbers in braces on next line

\maketitle 

\section{Introduction}

In a recent paper \cite{Kurian}, the dipole structure of DNA has been studied by examining the molecular dipole components induced due to London dispersion forces between base pairs. In particular, delocalized electrons in the base pairs of DNA were analyzed and theoretically shown to stimulate dipole formation at the molecular level. These induced dipoles generate collective modes that are suitably fine-tuned for enzymatic activity resulting in double-strand breaks in the DNA helix. This class of enzymes rapidly scans the DNA  \cite{pingoud2005type} searching for target recognition sequences, which exhibit a marked palindromic mirror symmetry between one DNA strand and its complement. Specific binding of target sequence produces conformational changes in the enzyme and DNA, with characteristic release of water and charge-countering ions from the DNA-protein interface. Under optimum biological conditions, concerted cutting of both strands then occurs without producing intermediate single-strand cuts \cite{stahl1998intra, stahl1996introduction, halford1988modes, maxwell1982salgi}, which requires, in order to occur, synchronization of dipole vibrational modes between spatially separated nucleotides and enzymatic molecular subunits.  The dipolar correlations between DNA and enzyme are required in order for these enzymes to cleave DNA in a manner that preserves the palindromic symmetry of the double-stranded substrates to which they bind.  

\begin{figure*}
	\begin{subfigure}[b]{0.5\linewidth}
		\includegraphics[width=\textwidth]{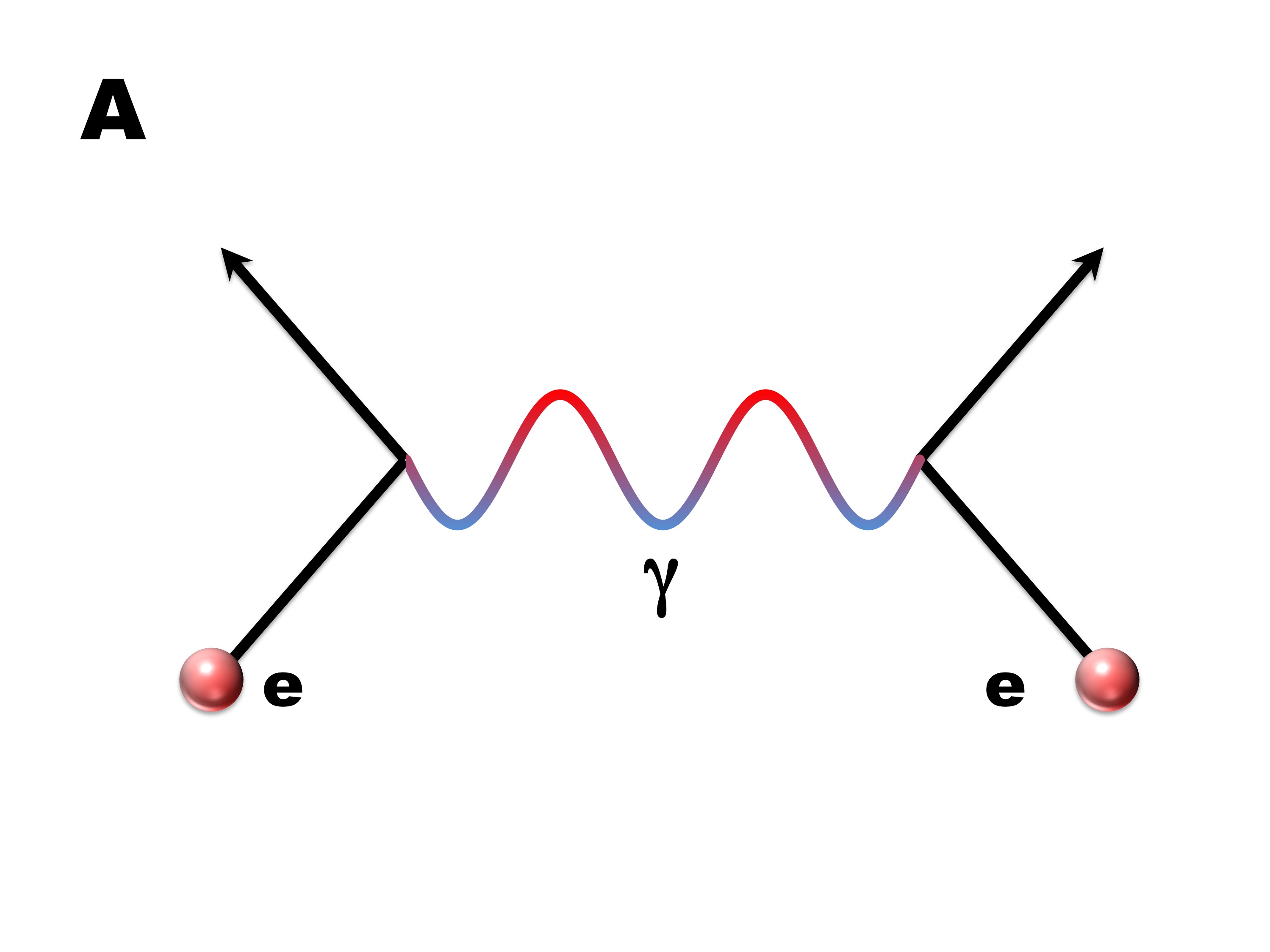}
		\label{subfigA}
	\end{subfigure}%
	\begin{subfigure}[b]{0.5\linewidth}
		\includegraphics[width=\textwidth]{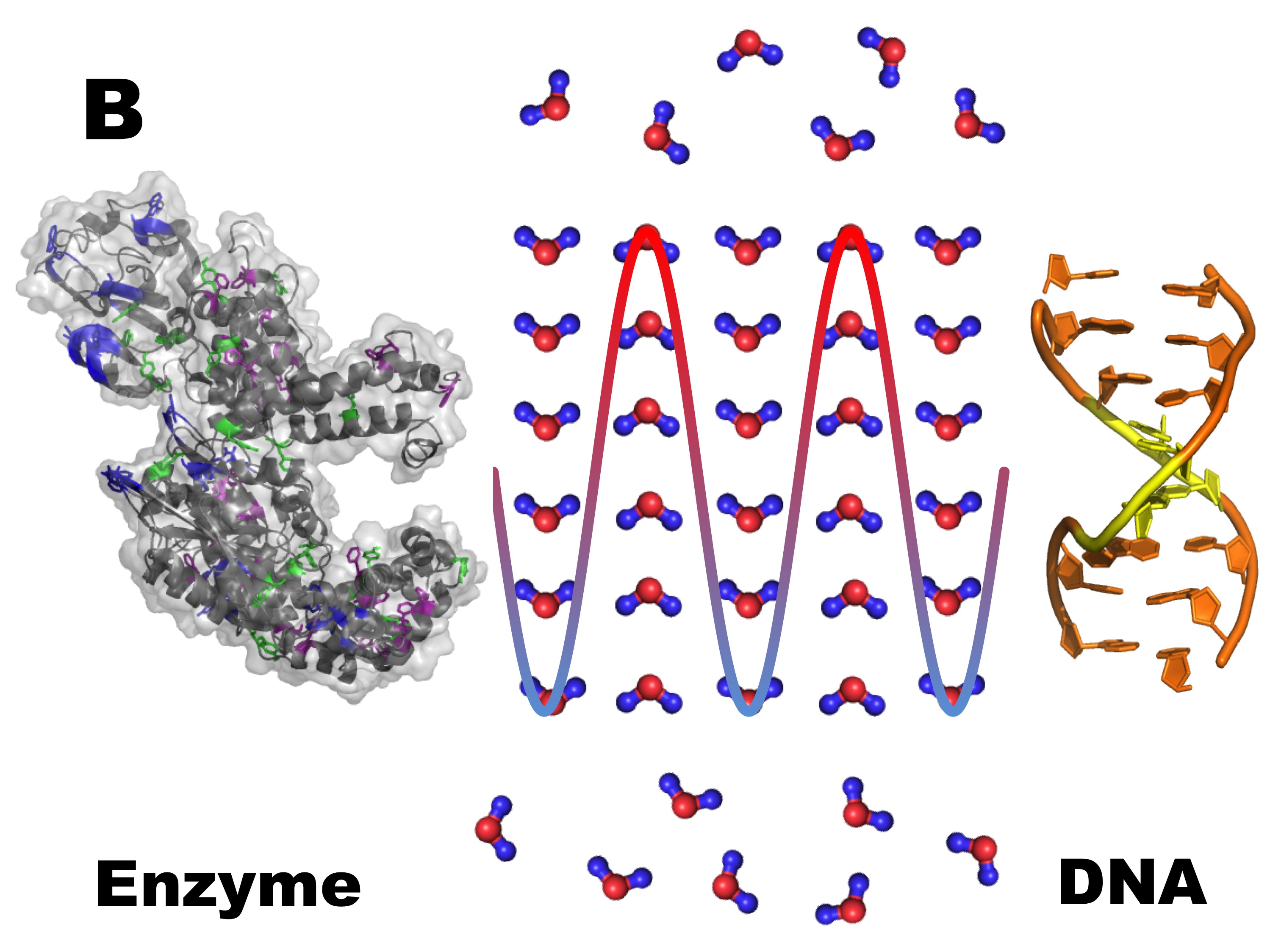}
		\label{subfigB} 
	\end{subfigure}
\captionsetup{justification=raggedright, singlelinecheck=false}
\caption{\textbf{Mediating wave fields or quanta in subatomic and biological physics}. (A) Electron-electron correlations are mediated by photons in quantum field theory. (B) Analogously, long-range correlations in the molecular water field between DNA and enzymes may be mediated by dipole waves. Note that these are schematic renderings, neither drawn to scale nor representative of the actual orientations of water molecules.}
\label{fig:feynman}
\end{figure*}

Although the evolution from sequence recognition to catalysis is perhaps the least understood aspect of the enzymology, the synchronous long-range correlations over distance are a hallmark of collective molecular dynamics \cite{Frohlich, Preto}. However, the study of the chemistry of DNA-enzyme interactions, and perhaps the whole range of biomacromolecular interactions, is still fraught with the lacuna in our understanding between the intrinsically stochastic molecular kinematics and the high efficiency and precise targeting of enzymatic catalytic activity \cite{narrowescape, Cosic}. This highly efficient, ultra-precise coordination has been shown to confound the explanatory reach of statistical methods for describing average regularities in bulk matter \cite{SchroedingerWhatisLife}. What is needed is of course not in opposition to the current state of knowledge derived from biochemistry. On the contrary, what we propose to shed light on is the quantum dynamical basis of the molecular interactions so as to provide a solid physical foundation for the biochemistry results. Such inquiry develops in many respects along the same path of the traditional study of the molecular dipole and multipole dynamical structure of the electronic quantum conformational shells determining the properties of interacting molecules in chemistry.

By following such a line of thought, our aim in the present paper is to deepen the understanding of DNA-enzyme interactions by characterizing quantitatively the nature of long-range correlations between DNA and enzyme in water. Water is the matrix of all known living systems and constitutes about 65\% of the human body by mass and about 99\% by number of molecules. On the one hand, we exploit the basic feature of any statistically oriented analysis, namely the fact that biomolecules are not isolated from their  environment of water molecules. Thus, we have to face the complexity arising from dealing with a large number of molecular components and the quantum vibrational modes of their dipoles, which brings us to consider the quantum field theory (QFT) formalism. On the other hand, the generation of collective mode dynamics derived in the framework of spontaneously broken gauge symmetry theories can explain the observation of highly ordered patterns characterizing the catalytic activity, in space and in time (time ordered sequences of steps in the chemical activity). In a rather natural way we are thus led to adopt in the study of DNA-enzyme interactions the QFT paradigm of gauge theories, namely that the interaction between two systems is realized through the exchange between them of a mediating wave field or quantum, in analogy with the photon exchanged by two electric charges in quantum electrodynamics (see Figure \ref{fig:feynman}). Due to the relevant role played by electric dipoles in the molecular interactions under study, we center our investigation on a generalized model of the radiative dipole wave field mediating the molecular interactions between DNA and enzyme in water. This model is discussed in Sections III and IV.

The paper is organized as follows. In Section II we discuss and partially review the molecular dipole structure of DNA and enzyme. To connect with previous work, we consider the case of the type II restriction endonuclease \textit{Eco}RI and, for generality, we similarly analyze the \textit{Taq} DNA polymerase, which is widely used in polymerase chain reaction (PCR) processes for the amplification of DNA sequences. In Section III we study the collective dipole dynamics of water molecules in the presence of the  DNA radiative dipole field. In Section IV we examine how the water dipole wave field interacts with the enzyme radiative dipole field. The water dipole field is identified as the mediating wave field in the DNA-enzyme interaction. Concluding remarks are presented in Section V. The appendices A, B, and C report some details on the enzyme systems, mathematical formalism, and analysis of the system ground states, respectively.

\section{Induced Dipole Networks in DNA Sequence and Enzyme Systems}

In this section we present the computation of the collective dipole behavior in the DNA and enzyme molecules by considering the interactions between their constituent aromatic rings. These rings, present in both DNA base pairs and enzyme amino acids as well as a host of other biomolecules, contain conjugated planar ring systems with delocalized $\pi$ electrons shared across the structure, instead of permanent, alternating single and double bonds. Benzene is the canonical example of such a ring. This confers on aromatic compounds an unusual stability and low reactivity, but also provides an ideal structure for the formation of electric dipoles, which can interact to produce electromagnetic couplings in biology.

\subsection{DNA Sequence}
We consider first the linearized DNA sequence, with polarizability data for its four bases given in Table \ref{DNAalpha}. The symmetry and regularity of the molecule about its helical axis simplifies the calculation relative to the one for enzyme aromatic networks. By resorting to the results of Ref. \cite{Kurian}, we observe that in DNA the delocalized electrons belong to the planar-stacked base pairs that serve as ``ladder rungs'' stepping up the longitudinal helix axis. The induced dipole formation is generated by Coulombic interactions between electron clouds. The helix rungs can be visualized as electronically mobile sleeves vibrating with small perturbations around their fixed positively charged core.

\begin{table}[!htbp]
\centering
\captionsetup{justification=raggedright, singlelinecheck=false}
\caption{Polarizabilities for DNA bases \cite{mcweeny1962perturbation, papadopoulos1988polarisability, basch1989electrical}, in units of 1 au = $1.64878 \times 10^{-41} \text{C}^2 \text{m}^2 \text{J}^{-1}$.} 
\label{DNAalpha}
\begin{tabular}{ c  c  c  c}
\hline
\hline
DNA base & $\alpha_{xx}$ & $\alpha_{yy}$ & $\alpha_{zz}$  \\ \hline
Adenine (A) & 102.5  & 114.0 & 49.6 \\
Cytosine (C) & 78.8  & 107.1 & 44.2 \\
Guanine (G)  & 108.7 & 124.8 & 51.2\\
Thymine (T) & 80.7    & 101.7  & 45.9\\
\hline \hline
\end{tabular}
\end{table}

Consider a molecule of length $N$ nucleotides. The Hamiltonian is
\begin{eqnarray} \label{eq:Hamiltonian}
\nonumber H_{DNA}=\sum^{N-1}_{s=0} \frac{\mathbf{p}_s^2}{2m_s} &+ \frac{m_s}{2}\left(\omega_{s,xx\,}^{2}x_s^2 +\omega_{s,yy\,}^2y_s^2 + \omega_{s,zz\,}^2z_s^2\right ) \\
&+ V_s^{int},
\end{eqnarray}
where $\mathbf{r}_s=(x_s, y_s, z_s)$ are the displacement coordinates between each electron cloud and its base-pair core, the coordinates $x_s, y_s$ span the transverse plane of each base-pair cloud, and the $z_s$ are aligned along the longitudinal axis. The dipole-dipole interaction terms are given by
\begin{equation} \label{eq:dip-dip}
V_s^{int} = \frac{1}{4\pi\epsilon_0 d^3}\left[\gv{\pi}_s \cdot \gv{\pi}_{s+1}- \frac{3(\gv{\pi}_s \cdot \mathbf{d})(\gv{\pi}_{s+1} \cdot \mathbf{d})}{d^2}\right],
\end{equation}
with $\mathbf{d} = d\hat{\mathbf{z}}$ connecting the centers of nearest-neighbor base-pair dipoles $\gv{\pi}_s = Q\mathbf{r}_s$ and $\gv{\pi}_{s+1} = Q\mathbf{r}_{s+1}$.

As shown in Table \ref{DNAbpomega}, $\omega_{s,ii\,}$ are the diagonal elements of the angular frequency tensor for each base-pair electronic oscillator and are determined from polarizability data:\begin{equation} \label{eq:omegatensor}
\omega_{A:T, \,ii} = \sqrt{\frac{Q^2}{m(\alpha_{A, \,ii}+\alpha_{T,\, ii})}},
\end{equation}
and similarly for $\omega_{C:G, \,ii}$. As shown in Appendix B, Eq.~(\ref{eq:omegatensor}) may be derived from the fundamental dipole relation $\gv{\pi}=\gv{\alpha}\cdot \mathbf{E}$. The numerical tensor elements $\alpha_{ii}$ account for anisotropies, which have been determined from perturbation theory \cite{mcweeny1962perturbation}, simulation \cite{papadopoulos1988polarisability}, and experiment \cite{basch1989electrical} generally to within five percent agreement.  The mass and charge of an electron are used because single electrons would be entangled through the base-pair couplings.

\begin{table}[!htbp]
\centering
\captionsetup{justification=raggedright, singlelinecheck=false}
\caption{Electronic angular frequencies calculated from polarizability data for DNA base pairs (bp),  in units of $10^{15}$ radians per second.}
\label{DNAbpomega}
\begin{tabular}{ c  c  c  c }
\hline
\hline
bp & $\omega_{xx}$ & $\omega_{yy}$ & $\omega_{zz}$ \\ \hline
A:T & 3.062 & 2.822 & 4.242\\
C:G & 3.027 & 2.722 & 4.244\\ \hline \hline
\end{tabular}
\end{table}

Due to the twist in the helix about the longitudinal axis, we must account for cross terms between directional components of two interacting dipoles. Choosing a single coordinate frame that corresponds with $(x_0, y_0, z_0)$ of the $0$th base pair, and relating such coordinates to the $(x_s, y_s, z_s)$ ones, the interaction potential for the $s$th electronic oscillator can be written as
\begin{align} \label{eq:dipolepotential}
V_s^{int} = \frac{Q^2}{4\pi\epsilon_0d^3}[&x_s x_{s+1} \cos \theta + y_s y_{s+1} \cos \theta \\\nonumber
&+ (y_s x_{s+1} - x_s y_{s+1}) \sin \theta- 2z_s z_{s+1}],
\end{align}
where the orientation of the helix and the twist angle are reflected in the different factors for the quadratic couplings.

By introducing the normal-mode lowering operator
\begin{equation} \label{eq:annihilation}
a_{s, j}=\sqrt{\frac{m\Omega_{s,j}}{2\hbar}}(\v{r}^\prime_s)_j+ \frac{i}{\sqrt{2m \hbar \Omega_{s,j}}}(\v{p}^\prime_s)_j,
\end{equation}
and its conjugate raising operator $a_{s,j}^\dagger$ for the $s$th normal mode of the collective electronic oscillations for the $j=xy$ or $j=z$ potential,  with
\begin{align}
(\v{r}^\prime_s)_j = \sum \limits_{n=0}^{N-1} (\v{r}_n)_j \, \exp\left(-\frac{2\pi i ns}{N}\right), \\ \nonumber
(\v{p}^\prime_s)_j = \sum \limits_{n=0}^{N-1} (\v{p}_n)_j \, \exp \left(-\frac{2\pi i ns}{N}\right),
\end{align}
the Hamiltonian in Eq. (\ref{eq:Hamiltonian}) takes the standard diagonalized form
\begin{equation} \label{eq:numop}
H_{DNA}=\sum_j H_j= \sum_j \sum^{N-1}_{s=0}\hbar\Omega_{s,j} \left (a_{s,j}^\dagger a_{s,j} +\frac{1}{2} \right).
\end{equation}
The eigenstates of $H_j$ are given by
\begin{equation}
\ket{\psi_{s,j}}=a_{s, j}^\dagger \ket{0},
\end{equation}
where $s=0,1,\dots,N-1$ for the $j=xy, z$ potential. Only the lowest energy states will be considered because these modes are the most easily excited.

In order to obtain the collective eigenmode frequencies for the oscillations, by separating Eq.~(\ref{eq:Hamiltonian}) into energy contributions from transverse ($H_{xy}$) and longitudinal ($H_z$) modes, we may write the symmetric longitudinal potential matrix $\mathbf{V}_z$ for a four-bp sequence and the diagonal kinetic matrices $\mathbf{T}_j$ as given in Appendix B. The problem then consists of solving
\begin{equation}
\det(\mathbf{V}_j - \Omega_{s,j}^2\mathbf{T}_j) = 0
\label{eq:goldstein}
\end{equation}
for the eigenfrequencies $\Omega_{s,j}$.

When homogeneity is assumed in the sequence $(m_s=m, \omega_{s,zz}=\omega, \gamma_{s,s+1}^z = \gamma)$, the longitudinal mode frequencies in the four-bp case take on a beautifully simple form:
\begin{align} \label{eq:goldenrat}
\nonumber \Omega_{0,z}^{\,2} &= \omega^2-\varphi \frac{\gamma}{m}\\
\nonumber \Omega_{1,z}^{\,2} &= \omega ^2- (\varphi-1)\frac{\gamma }{m}\\
\nonumber \Omega_{2,z}^{\,2} &= \omega ^2+ (\varphi-1)\frac{\gamma }{m}\\
\Omega_{3,z}^{\,2} &= \omega ^2+ \varphi \frac{\gamma}{m},
\end{align}
where $\varphi = \left(1+\sqrt{5}\right)/2$ is the golden ratio. For more details, see Ref.~\cite{Kurian}.

Of course, DNA has no observed electronic transitions---meaning absorption spectra of individual chromophores leading to electronic excitation and fluorescence---in the 0.1 - 1.7 eV range calculated in Table \ref{tab:EcoRI}. However, the type of processes we describe here are coherent oscillations that arise from induced dipole-dipole correlations between constituent aromatic rings. These dipolar correlations do not require photoabsorption-induced electronic excitation. This coherent behavior stems from London dispersion forces between aromatic rings and has been described in the literature \cite{Ambrosetti, Frohlich2}. 

\begin{table}[!htbp]
\centering
\captionsetup{justification=raggedright, singlelinecheck=false}
\caption{\textit{Eco}RI DNA recognition sequence (GAATTC) zero-point modes, in units of $\varepsilon_{P-O} \simeq 0.23$ eV, as a function of inter-base-pair spacing $d$. The helix twist angle $\theta \simeq \pi/5$ is constant for the six-bp sequence.}
\begin{tabular}{ ccccc }
\hline \hline
$d$ (\r{A}) & $\hbar\Omega_{0,xy}/2$ & $\hbar\Omega_{1,xy}/2$ & $\hbar\Omega_{0,z}/2$ & $\hbar\Omega_{1,z}/2$ \\ \hline
        $3.0$ &   2.86       &  3.07     &    4.45  & 7.35 \\
        $3.2$ &   3.11       &  3.31      &   4.78  & 7.14 \\
        $3.4$ &   0.53       & 1.35       &   1.99  & 5.02 \\
        $3.6$ &   1.70       & 2.11       &   3.02  & 5.20 \\
        $3.8$ &   0.40       & 0.82        &  1.67  & 3.65 \\  \hline \hline
\end{tabular}
\label{tab:EcoRI}
\end{table}

We remark that the energy required \textit{in vivo} to break a single phosphodiester bond in the DNA helix backbone is $\varepsilon_{P-O} \simeq 0.23$ eV \cite{dickson2000determination}, which is less than two percent of the energy required to ionize the hydrogen atom, but about ten times the physiological thermal energy ($k_B T$). This value of $\varepsilon_{P-O}$ is comparable with the quantum of biological energy released during nucleotide triphosphate (e.g., ATP) hydrolysis; it ensures that the bonds of the DNA backbone are not so tight as to be unmodifiable, but remain strong enough to resist thermal degradation. As shown in Table \ref{tab:EcoRI}, at the standard inter-base-pair spacing of $3.4\,$\r{A}, we calculate the ground state longitudinal oscillation (a zero-point mode) to within $0.5\%$ of $2\varepsilon_{P-O}$, the energy required for double-strand breakage.

It is also remarkable that for \textit{Eco}RI, considered in more detail below, the difference in free energy between the nonspecific and specific complex (i.e., clamping energy) is approximately $2\varepsilon_{P-O}$. Thus, enzyme clamping imparts the quanta of energy necessary to excite the longitudinal mode. Because it is a collective, normal-mode oscillation (where components vibrate in synchrony), this lowest-energy mode along the main axis of the DNA sequence coherently transports the quanta to break two phosphodiester bonds simultaneously. Parametrization of this DNA model is briefly discussed in Appendix B.

\subsection{Enzyme Systems}

\begin{figure*}[htb]
\centering
\includegraphics[width=0.9\textwidth]{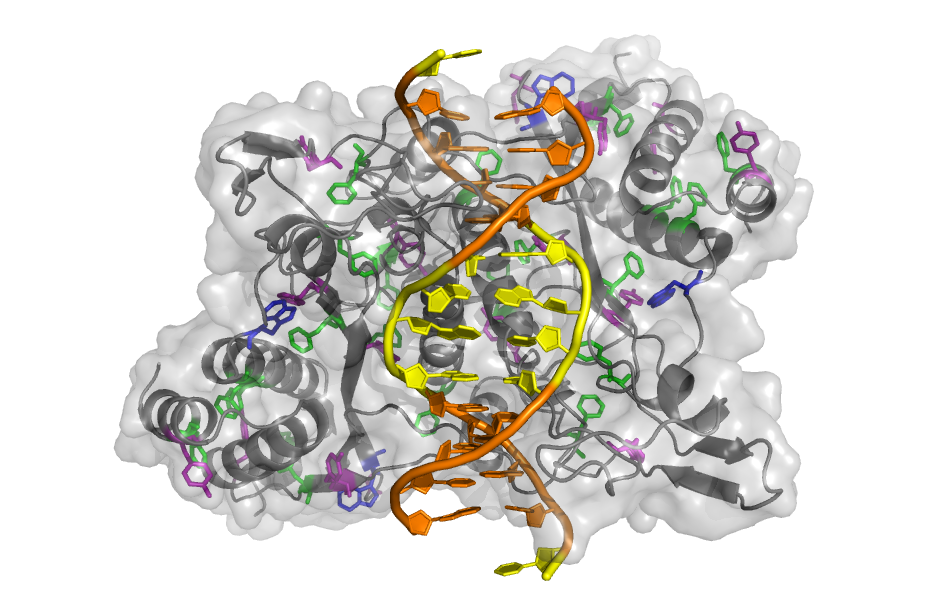}
\captionsetup{justification=raggedright, singlelinecheck=false}
\caption{\textbf{Aromatic amino acid network in \textit{Eco}RI}. Tryptophan (blue), tyrosine (purple), and phenylalanine (green) form induced dipole networks in \textit{Eco}RI, shown here bound to its double-stranded DNA substrate, with A:T (yellow) and C:G (orange) base pairs highlighted. Other amino acids (gray) are displayed in the context of their secondary structures within the enzyme. Image of \textit{Eco}RI (PDB ID: 1CKQ) at 1.85 \r{A} resolution created with PyMOL.}
\label{fig:aroEco}
\end{figure*}

\begin{figure*}[htb]
\centering
\includegraphics[width=0.9\textwidth]{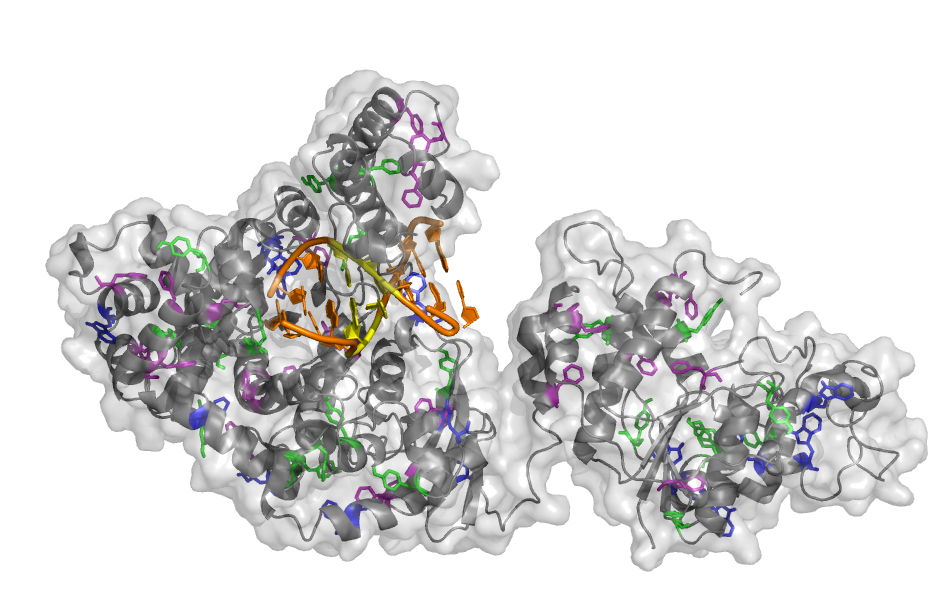}
\captionsetup{justification=raggedright, singlelinecheck=false}
\caption{\textbf{Aromatic amino acid network in \textit{Taq} DNA polymerase}. Tryptophan (blue), tyrosine (purple), and phenylalanine (green) form induced dipole networks in \textit{Taq} polymerase, shown here bound to a blunt-ended duplex DNA in the polymerase active-site cleft, with A:T (yellow) and C:G (orange) base pairs highlighted. Other amino acids (gray) are displayed in the context of their secondary structures within the enzyme. Image of \textit{Taq} polymerase (PDB ID: 1TAU) at 3.0 \r{A} resolution created with PyMOL.}
\label{fig:aroTaq}
\end{figure*}

We will consider two specific enzymes---\textit{Eco}RI and \textit{Taq} polymerase---that represent two broad enzyme classes used in DNA metabolism, namely type II restriction endonucleases and DNA polymerases, respectively. The former group initiates DNA double-strand breaks with high specificity and the latter replicates DNA strands with high efficiency.

We will first consider the enzyme \textit{Eco}RI, shown in Figure \ref{fig:aroEco} and the prototype for a class of enzymes called type II restriction endonucleases that recognize very specific DNA sequences. \textit{Eco}RI was the first of this class to be discovered, found in \textit{E.coli} bacteria (see Appendix A). It initiates a double-strand break (DSB) in the DNA helix after recognizing the sequence GAATTC. The mechanism of this DSB synchronization has been the subject of much investigation and was recently proposed to be correlated through quantum entanglement \cite{Kurian}. One interesting characteristic of the \textit{Eco}RI recognition site is the palindromic symmetry of the sequence. By the rules of DNA base pairing, the complementary DNA sequence is CTTAAG, so GAATTC is the mirror image of its complement.

The other enzyme we consider is the \textit{Taq} DNA polymerase, shown in Figure \ref{fig:aroTaq} and used widely in molecular biology labs throughout the world for rapid amplification of DNA sequences. While \textit{Taq} does not recognize specific sequences like \textit{Eco}RI, it employs the use of an interesting DNA clamp. The DNA clamp fold is an accessory protein that assembles into a multimeric structure completely encircling the DNA double helix as the polymerase adds nucleotides to the growing strand. By preventing dissociation of the enzyme from the template DNA strand, the clamp dramatically increases the number of nucleotides the polymerase can add to the growing strand per binding event, up to 1,000-fold. The clamp assembles on the DNA at the replication fork, where the double helix effectively ``unzips'' into single strands, and ``slides'' along the DNA with the advancing polymerase, aided by a layer of water molecules in the central pore of the clamp between the DNA and the protein surface. Because of the toroidal shape of the assembled multimer, the clamp cannot dissociate from the template strand without also dissociating into monomers.

We are particularly interested in the polarizability of the amino acids, which are the building blocks of protein and enzyme systems. The most polarizable amino acids are termed \textit{aromatic}, and as with DNA we will examine the effect of induced dipole interactions between these constituent ring structures. See Tables \ref{Trpalpha} and \ref{Trpomega} for aromatic amino acid polarizabilities and electronic angular frequencies, respectively.

\begin{table}[!htbp]
\centering
\captionsetup{justification=raggedright, singlelinecheck=false}
\caption{Polarizabilities for aromatic amino acids \cite{indole, phenol, benzene1, benzene2, benzene3}, in units of 1 au = $1.64878 \times 10^{-41} \text{C}^2 \text{m}^2 \text{J}^{-1}$, with $\overline{\alpha} = \sqrt{\alpha_{xx}^2 + \alpha_{yy}^2 + \alpha_{zz}^2}$. For further details, see comments surrounding Eq.~(\ref{eq:alphas}).} 
\label{Trpalpha}
\begin{tabular}{ ccccc }
\hline
\hline
Amino Acid & $\overline{\alpha}$ & $\alpha_{xx}$ & $\alpha_{yy}$ & $\alpha_{zz}$ \\ \hline
Trp & 153.4 & 119.5 & 91.6 & 29.4 \\
Tyr & 129.3  & 89.5 & 43.0 & 82.9 \\
Phe &118.1 & 79.0 & 79.0 & 38.6 \\
\hline \hline
\end{tabular}
\end{table}

Interestingly, the most aromatic amino acids (tryptophan, tyrosine, and phenylalanine) are also the most hydrophobic. Various proposals have been put forward arguing the biological significance of such intra-protein hydrophobic pockets. London force dipoles in such regions could couple together and oscillate coherently, thus generating a radiative field.

\begin{table}[!htbp]
\centering
\captionsetup{justification=raggedright, singlelinecheck=false}
\caption{Electronic angular frequencies calculated from polarizability data for aromatic amino acids, in units of $10^{15}$ radians per second.}
\label{Trpomega}
\begin{tabular}{ c c c c c }
\hline
\hline
Amino Acid & $\overline{\omega}$ & $\omega_{xx}$ & $\omega_{yy}$ & $\omega_{zz}$   \\ \hline
Trp & 3.338 & 3.782 & 4.320 & 7.622 \\
Tyr & 3.635 & 4.370 & 6.305 & 4.541\\
Phe & 3.803 & 4.653 & 4.653 & 6.654 \\
 \hline \hline
\end{tabular}
\end{table}

In regards to the spacing between aromatic amino acids in protein, when considering tryptophan, tyrosine, and phenylalanine, the spacings are as close as 5 \r{A} in the enzymes of interest, comparable to the inter-bp spacing of DNA (3.4 \r{A}).  The use of the point-dipole approximation, rather than an extended dipole description, can be questioned due to these close separations, but this is a first-order estimate of the contribution of these effects. Additionally, because we are considering interactions between all aromatic rings, the point-dipole approximation would indeed hold for the longer distances, which are the majority. 

As shown in Table \ref{Trpomega}, $\omega_{s,ii\,}$ are the diagonal elements of the angular frequency tensor for each amino-acid dipole and are determined from the polarizability data:
\begin{equation} \label{eq:aatensor}
\omega_{aa, \,ii} = \sqrt{\frac{Q^2}{m\alpha_{aa, \,ii}}}.
\end{equation}
The mass and charge of an electron are used because the amino acids are approximated as charge-separated dipoles. As shown in Appendix B, Eq.~(\ref{eq:aatensor}), like Eq.~(\ref{eq:omegatensor}),  may be derived from the fundamental dipole relation $\gv{\pi}=\gv{\alpha}\cdot \mathbf{E}$. 

\begin{figure*}[htb]
\includegraphics[width=0.9\textwidth]{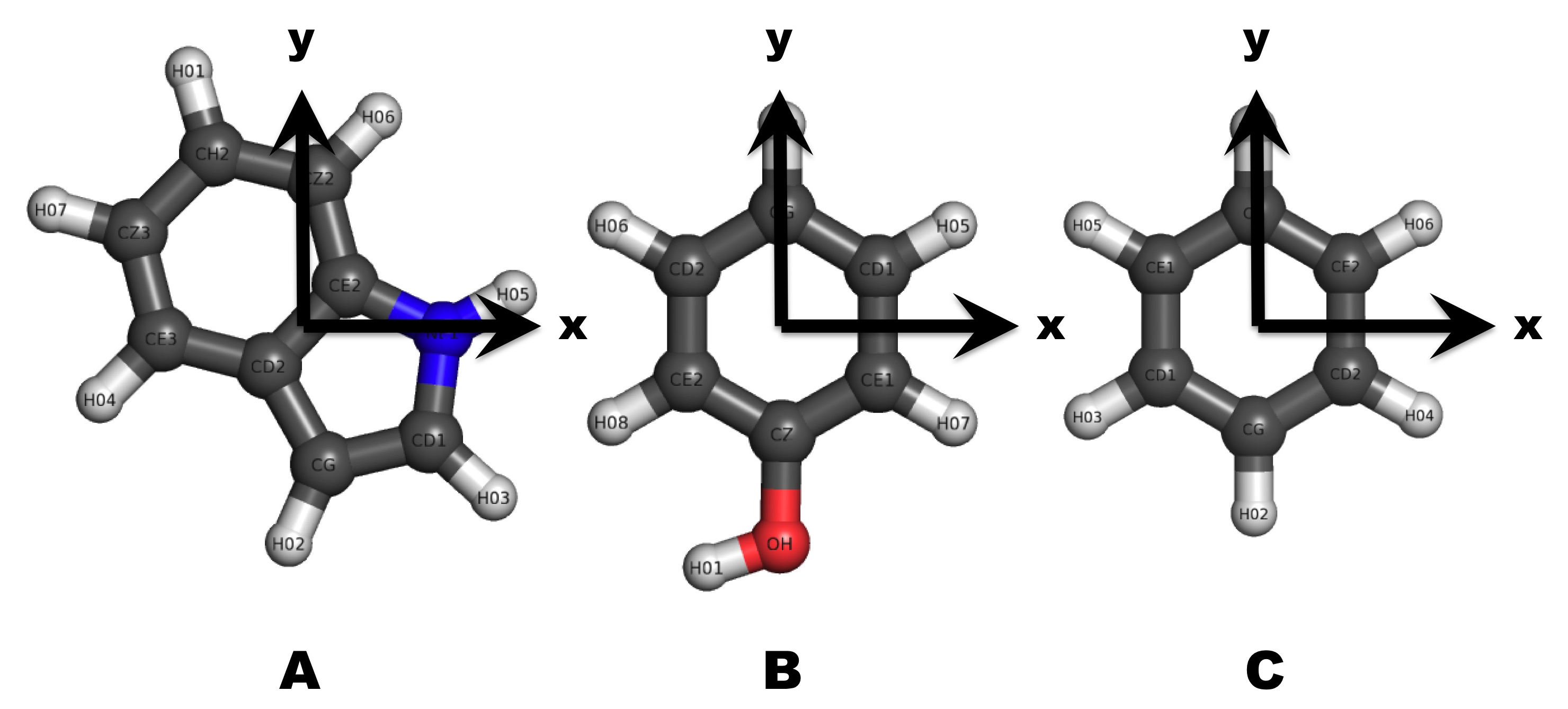}
\captionsetup{justification=raggedright, singlelinecheck=false}
\caption{\textbf{Orientation geometries and molecular centers of aromatic ring structures in amino acids}. (A) Indole ring in tryptophan. (B) Phenol ring in tyrosine. (C) Benzene ring in phenylalanine. The positive $z$-axis is directed out of the page.}
\label{indole}
\end{figure*}

Here we use the values and principal directions of the polarizability tensors based on the experimental and theoretical calculations for indole \cite{indole}, phenol \cite{phenol}, and benzene \cite{benzene1, benzene2, benzene3}, to represent the polarizability tensors of the amino acids tryptophan, tyrosine, and phenylalanine, respectively, because each ring structure defines the aromaticity of the corresponding amino acid in which it is contained. Figure \ref{indole} illustrates the orientation geometries of the principal directions and the molecular centers used in calculating the dipole displacement coordinates between aromatic amino acids in the aromatic coordinate frame. We use only the diagonal elements of the polarizability tensor $\al_{xx}, \al_{yy}, \al_{zz},$ and neglect off-diagonal terms. After alignment of the polarizabilities with the orientation of the aromatic amino acids in the protein coordinate space, we take the magnitude of the average polarizability to be
\be \lab{eq:alphas}
\overline{\al} = \sqrt{\al_{xx}^{'2}+\al_{yy}^{'2} + \al_{zz}^{'2}}~,
\ee
where the induced dipole direction for each aromatic is defined by the vector $(\alpha_{xx}', \alpha_{yy}', \alpha_{zz}')$ in the protein coordinate frame.

As we did for the DNA molecule in the first part of the present section, we proceed with the Hamiltonian for such a collection of $N$ aromatic amino acids:
\begin{equation} \label{eq:EnzHamiltonian}
H_{enz}=\sum^{N-1}_{s \neq t}\frac{\mathbf{p}_s^2}{2m_s} + \frac{m_s \overline{\omega}_{s}^{2}}{2}\left(x_s^2 +y_s^2+z_s^2\right ) + V_{s,t}^{int},
\end{equation}
where the displacement coordinates $\mathbf{r}_s=(x_s, y_s, z_s)$ are defined as previously done between each electron cloud and its amino acid core. Here $\overline{\omega}_{s\,}$ are derived from the polarizabilities $\overline{\alpha}_s$ in Table \ref{Trpalpha} through Eq.~(\ref{eq:aatensor}), and the term $V_{s,t}^{int} = \frac{1}{4\pi\epsilon_0 d_{s,t}^3}\left[\gv{\pi}_s \cdot \gv{\pi}_{t}- \frac{3(\gv{\pi}_s \cdot \mathbf{d}_{s,t})(\gv{\pi}_{t} \cdot \mathbf{d}_{s,t})}{d_{s,t}^2}\right]$ describes the pairwise interactions between each amino acid electric dipole in the enzyme's aromatic network. The aromatic networks for \textit{Eco}RI and \textit{Taq} are shown in Figures \ref{fig:aroEco} and \ref{fig:aroTaq} and are determined from crystal structures with identification codes 1CKQ \cite{EcoCKQ} and 1TAU \cite{TaqEom}, respectively, in the Protein Data Bank \cite{PDB}. The 1CKQ PDB file must be duplicated appropriately to account for the symmetric homodimer formed from two identical protein subunits. Tryptophan, the most strongly polarizable amino acid, will generally contribute most to the collective electronic behavior of each enzyme.

We return to solving Eq.~(\ref{eq:goldstein}), but this time in the enzyme case for larger, denser potential matrices $V_{s,t}^{int}$. This largeness and denseness is due to the sheer number of aromatic amino acids and the pairwise coupling interactions between them. We simplify this task by recognizing that Eq.~(\ref{eq:goldstein}) is similar to the characteristic equation for matrix eigenvalues:
\begin{equation}  \label{eq:eigenvalue}
det (\mathbf{A} - \lambda_s \mathbf{I})=0,
\end{equation}
with $\mathbf{V}$ in place of $\mathbf{A}$ and $\lambda_s = m_e\Omega_s^2$. Numerical packages can then be used to quickly and efficiently solve for the eigenvalues of $\mathbf{V}$.

\begin{figure*}
	\begin{subfigure}[b]{0.5\linewidth}
		\includegraphics[width=\textwidth]{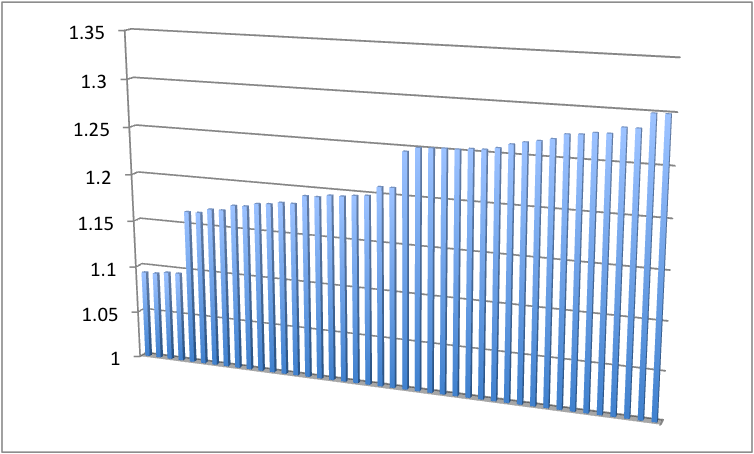}
		\subcaption{\textit{Eco}RI} \label{fig:1erimodes}
	\end{subfigure}%
	\begin{subfigure}[b]{0.5\linewidth}
		\includegraphics[width=\textwidth]{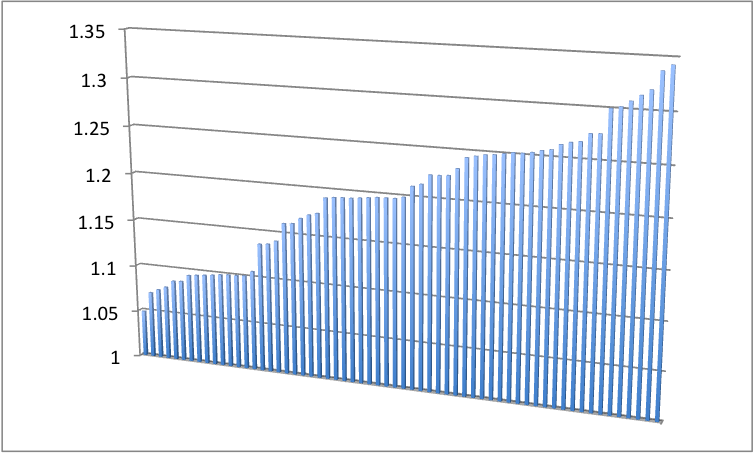}
		\subcaption{\textit{Taq}} \label{fig:1taumodes} 
	\end{subfigure}
\captionsetup{justification=raggedright, singlelinecheck=false}
\caption{\textbf{Collective dipole oscillations in aromatic amino acid networks of DNA-interacting enzymes.} Collective normal-mode solutions to networks of aromatic induced dipoles in \textit{Eco}RI (\ref{fig:1erimodes}) and \textit{Taq} (\ref{fig:1taumodes}) are within the energy range of the collective dipole modes of DNA bounded by its relevant protein clamps. These states are precisely those for which the number operator $n_{s} = a_{s}^\dagger a_{s}$ acts in the diagonalized form of Eq. (\ref{eq:EnzHamiltonian}), analogous to Eq. (\ref{eq:numop}) for DNA, to produce a zero eigenvalue, thus identifying them with zero-point modes that exist independent of external excitations and arise from the ground state of the enzyme aromatic network. Data for the collective modes (arranged along the abscissa axis according to increasing energy) are presented in units of eV on the ordinate axis.}
\label{enzcohmodes}
\end{figure*}

Collective modes for \textit{Eco}RI and \textit{Taq} are presented in Figure \ref{enzcohmodes}. The average energy for the \textit{Eco}RI modes is 1.22$\varepsilon_{P-O}$ and for the \textit{Taq} modes is 1.20$\varepsilon_{P-O}$. Note that \textit{Taq} is a larger enzyme and has more dipolar vibrational modes than \textit{Eco}RI (61 vs. 42) because these modes correspond to the number of aromatic amino acids in each enzyme. Though proteins with aromatic amino acids typically absorb in the UV, around $4-5$ eV, this observation does not discount coherent oscillations of induced dipoles at longer wavelengths in protein aromatic networks due to the dipolar correlations we describe in this paper. 

In each enzyme, there appear to be four ``shelves'' of states, generally clustering near 1.10, 1.20, 1.25, and 1.30$\varepsilon_{P-O}$ (2041, 2226, 2319, and 2412 cm$^{-1}$). Carboxylic acids ubiquitous in proteins tend to form strongly hydrogen bonded dimers, which shift water's O-H stretching frequencies to between 2000-3500 cm$^{-1}$ depending on the strength of the hydrogen bonds \cite{IRSpec}. It is interesting and significant to our studies that the enzyme collective dipole modes displayed in Figure \ref{enzcohmodes} fall within the spectrum of DNA collective modes presented in Table \ref{tab:EcoRI} and described in full detail in Ref. \cite{Kurian}. The enzyme collective modes, like the DNA modes in Eq.~(\ref{eq:numop}), are dipole fluctuation modes and can be used to construct the creation and annihilation operators (see Eq.~(\ref{eq:annihilation})) for the radiative electromagnetic field quanta (i.e., photons) acting on the water molecules.

We thus arrive at a picture for the case of enzymes similar from the physical point of view to the one obtained above for DNA. The DNA and the enzyme molecules, although sharply different in their chemical structures, are  macromolecular systems physically characterized by their dipolar structure. According to our quantum electrodynamics description, dipole vibrational modes are the source of radiative electromagnetic (em) dipole fields through which the interaction between DNA and enzyme develops. Since they are embedded in the water environment whose molecules are also characterized by their molecular dipoles, DNA and enzyme ``talk'' to each other by means of their radiative dipole fields propagating in the water medium. This long-range communication is mediated by the water dipoles. The chemical activity between DNA and enzyme develops in and is conditioned by such a physical scenario. In order to study the physical dynamics providing the environmental conditions under which the chemical activity develops, we need to deepen the understanding of collective dipole modes of liquid water. This is done in the following section.

\section{The Water Dipole Field}

Water constitutes the environment in which  DNA and  enzyme macromolecules described in the previous sections are embedded. These macromolecules and the water molecules are endowed with electrical dipoles, whose quantum vibrational modes characterize their reciprocal interactions.
The picture emerging from the different interaction couplings, frequencies, amplitudes, and phase modulations in different space and time regions of the water environment provides the em imaging or em field pattern of the DNA-water-enzyme interaction. We will focus on the local properties of such a pattern of fields, thus dealing with densities rather than with physical quantities integrated over the whole volume, although integrations over some relatively small volumes will also be considered.

We study first the interaction between DNA and water. After that we will introduce the interaction with enzyme macromolecules. We will neglect the static dipole-dipole interactions between DNA and water dipoles and focus our attention instead on the radiative quantum electromagnetic (em) interaction.

Consider a system of molecules of liquid water, assumed to be homogeneous in a region whose linear  size $\ell$ will be elaborated in the course of our discussion (see the comments following Eq.~(\ref{9})). The magnitude of the water molecule dipole vector $\mathbf{d}_e$ is  given by $|\mathbf{d}_e | =2\,e\,d_{e}^{2}$, with  $d_{e} \simeq 0.2\,$\r{A} \cite{Franks} and $e$ denoting the electron charge. In the following discussion we use natural units, where $\hbar = 1 = c$.

The system is assumed to be at a non-vanishing temperature $T$. Under such conditions, in the absence of other actions, the system of water molecules is invariant under dipole rotations because the molecular dipoles are in general arbitrarily oriented and thus there is no preferred orientation direction. Our aim is to study the radiative dipole transitions in the water molecules induced by interactions with the DNA em field and all molecular water dipoles.

By closely following Ref. \cite{PRA2006} in our presentation, we denote by $\phi ({\bf x}, t)$ the complex dipole field collectively describing the $N$ molecule system in the unit volume $V$ so that integration over the sphere of radius $\bf r$  gives:
\begin{equation} \label{3} \int d\Omega |\phi ({\bf x},t)|^{2} = N~, \end{equation}
where $d\Omega = \sin \theta d \theta d \phi$ is the element of solid
angle and the polar coordinates  are denoted  by $(r, \theta, \phi)$.
It is convenient to introduce the rescaled field $\chi ({\bf x},t) =
\frac{1}{\sqrt{N}} \phi ({\bf x},t)$. The integration over the unit sphere then gives
\begin{equation} \label{4}
 \int d\Omega |\chi ({\bf x},t)|^{2} = 1~.  \end{equation}

Since the molecule density is assumed to be spatially uniform in the unit
sphere, the angular variables are the
only relevant ones. This allows to us to expand the field $\chi ({\bf x},t)$  in terms of spherical harmonics in the unit
sphere:
\begin{equation} \label{4a} \chi ({\bf x},t) = \sum_{l,m}
\alpha_{l,m}(t)Y^{m}_{l}(\theta, \phi) ~. \end{equation}

By setting $\alpha_{l,m}(t) = 0$ for $l \neq 0,~ 1$, Eq.~(\ref{4a}) reduces
to the expansion in the four levels  $(l,m) = (0,0)$ and $(1,m)$ for $m
= 0, \pm 1$. The populations of  these levels at thermal equilibrium, in the absence of
interaction, follow the Boltzmann distribution and are given by $N
|\alpha_{l,m}(t)|^{2}$. We will show that the populations of the $l=0$ and $l=1$ levels change due to the presence of DNA and enzyme. The amplitude of $\alpha_{1,m}(t)$ does
not depend on $m$ since, due to the dipole rotational invariance,  there is no preferred
direction in the dipole orientation. In other words, the time average of the polarization $P_{{\bf
n}}$ along any direction ${\bf n}$ must vanish. In order to show that this is indeed the case, let us introduce the notation
\bea \nonumber \al_{0,0}(t) &\equiv& a_{0}(t) \equiv
A_{0}(t)e^{i\de_{0}(t)},\\
 \al_{1,m}(t) &\equiv&  A_{1}(t)e^{i\de_{1,m}(t)}e^{-i\om_{0}t} \equiv a_{1,m}(t)
e^{-i\om_{0}t}, \lab{9} \eea
where $A_{0}(t)$, $A_{1}(t)$, $\de_{0}(t)$, and $\de_{1,m}(t)$ are real
quantities and  $\om_{0} \equiv
{1}/{I}$ (in natural units), where $I$ is the average moment of inertia of the water molecule.\footnote{The moment of inertia of water varies within a factor of three depending on the axis around which it is calculated: $1.0220 \times 10^{-40}$ g$\cdot$cm$^2$ for the axis in the plane of the water molecule with the origin on the oxygen and orthogonal to the H-O-H angle, i.e., parallel to the longest dimension of the molecule; $2.9376 \times 10^{-40}$ g$\cdot$cm$^2$ for the axis orthogonal to the plane of the water molecule with the origin on the oxygen; and $1.9187 \times 10^{-40}$ g$\cdot$cm$^2$ for the axis in the plane of the water molecule with the origin on the oxygen and bisecting the H-O-H angle (see Ref. \cite{Eisenberg}).} Note that $\om_{0}$ is also the eigenvalue of ${{\bf
L}^{2}}/{2I}$ on the state $(1,m)$, with ${\bf L}$ denoting the total angular momentum for the molecule. In fact, ${l(l+1)}/{2I} = {1}/{I}$. 

Our discussion will be restricted to the resonant radiative em modes, i.e., those for which
$\om_{0} \equiv k = {2\pi}/{\la}$, and we use the dipole approximation, where $\exp(i{\bf k}\cdot {\bf x}) \approx 1$, since we are interested in the collective behavior of the $N$ molecule system. The system linear size $\ell$ is thus constrained to be $\ell \leq \la = {2\pi}/{\om_{0}}$. We thus see that
$\om_{0}$ provides a relevant scale for the system. For water, ${2\pi}/{\om_{0}} \approx 1774 \,(\text{eV})^{-1} \approx 3.5 \times 10^{-2}\,$cm (in $\hbar = 1 =c$ units), taking $I = 1.95943\times 10^{-40}$ g$\cdot$cm$^2$ as the average of the moments of inertia presented in the footnote.

We now consider in an arbitrary reference frame the ${\bf z}$ axis and a vector ${\bf n}$ parallel to it. Use of the explicit expressions for the spherical harmonics
\bea \nonumber Y^{0}_{0} &=& \frac{1}{\sqrt{4\pi}}~,~~~ Y^{0}_{1}
=
\sqrt{\frac{3}{4\pi}}~ \cos~\theta~,\\
Y^{1}_{1} &=& - \sqrt{\frac{3}{8\pi}}~ \sin~\theta ~e^{i\phi} = - [Y^{-1}_{1}]^{*} ~, \lab{7}
 \eea
gives
\bea \nonumber P_{{\bf n}} &=& \int d\Om \chi^{*} ({\bf x},t)({\bf
x} \cdot {\bf n})\chi ({\bf x},t)\\ &=&
\frac{2}{\sqrt{3}}A_{0}(t)A_{1}(t)\cos(\om -\om_{0})t ~, \lab{8}
\eea
where $\om t \equiv \de_{1,0}(t) - \de_{0}(t)$. Eq. (\ref{8}) shows indeed that the time
average of $P_{{\bf n}}$ is zero, which in turn confirms that the three
levels $(1,m)$ for $m = 0, \pm 1$ are in the average equally
populated under normal conditions, thus $\sum_{m} ~|\al_{1,m}(t)|^{2} =  3 ~|a_{1}(t)|^{2}$, with $a_{1}(t) \equiv a_{1,m}(t)$ for any $m = 0,\pm1$.
The normalization condition (\ref{4}) then can be written as
\be \lab{4b} |\al_{0,0}(t)|^{2} + \sum_{m} |\al_{1,m}(t)|^{2} =
|a_{0}(t)|^{2} + 3 |a_{1}(t)|^{2}   = 1  \ee
at
any time $t$, so that, by putting  $Q \equiv |a_{0}(t)|^{2} + 3 ~|a_{1}(t)|^{2}$, we have
${\pa}Q/{\pa t} = 0$, which expresses the conservation of
the total number $N$ of molecules (see Eq. (\ref{3}) and the rescaling adopted for $\chi({\bf
x},t)$). Moreover, the relation
\be \lab{4c} \frac{\pa}{\pa t}{|a_{1}(t)|^{2}} = -
\frac{1}{3}~\frac{\pa}{\pa t}{|a_{0}(t)|^{2}} ~\ee
shows that, at each time $t$, the rate of change
of the population in each of the levels $(1,m)$ for $m = 0, \pm 1$,
equally contributes, in the average,  to the rate of change in the
population of the level $(0,0)$ (rotational invariance).

Eq.~(\ref{4b}) shows that the initial
conditions at $t = 0$ can be set as
\be \lab{4f} |a_{0}(0)|^{2} = \cos^{2} \theta_{0},~~
|a_{1}(0)|^{2} = \frac{1}{3}\sin^{2} \theta_{0}. %~~0 < \theta_{0} < \frac{\pi}{2}.
\ee
We exclude the values of $\theta_{0}$ corresponding to the physically unrealistic cases in which the state (0,0) is completely filled ($\theta_{0} = n\pi$, n integer) or completely empty ($\theta_{0} = (2n + 1){\pi}/{2}$, n integer). We will see that the lower bound for the parameter $\theta_{0}$ is imposed by the dynamics in a self-consistent way. Note that $\theta_{0} = {\pi}/{3}$ corresponds to the equipartition of the field modes among the four levels $(0,0)$ and $(1,m)$, as given by the
Boltzmann distribution when the temperature $T$ is much higher than the energy $E(k)$ of the field modes, $k_{B} T \gg E(k)$, which would be the case in our DNA-enzyme systems.

Let $u_{m}(t) \equiv U (t) e^{i\varphi_{m}(t)}$, with $U (t)$ and $\varphi_{m}(t)$ real quantities,  denote the field of the em modes generated by the DNA molecular dipoles (see Eqs. (\ref{eq:annihilation}), (\ref{eq:numop}), and the comments following Eq. (\ref{eq:eigenvalue})) and the ones generated by the molecular water dipoles. One also finds \cite{PRL1988,PRA2006} that the amplitude
$|u_{m}(t)|$ does not depend on $m$, so that we may write $|u(t)| \equiv |u_{m}(t)|$. Moreover, one can derive that the phases also do not depend on $m$; thus
$\varphi \equiv
\varphi_{m}$, $\de_{1}(t) \equiv \de_{1,m}(t)$.
We consider the coupling of $u(t)$ with water molecules in the transition $(1,m) \leftrightarrow (0,0)$.

The field equations for a water molecule in the field $u(t)$ are given in Refs. \cite{Knight,Heitler}:
\bea \nonumber i \frac{\partial \chi ({\bf x},t)}{\partial t} &=&
\frac{{\bf L}^{2}}{2I} ~\chi ({\bf x},t) - i \sum_{{\bf k}, r}
d_{e}\sqrt{\rho }~\sqrt{\frac{k}{2}} ~ ({\bf \ep}_{r} \cdot {\bf
x})\\
\nonumber &~&[u_{r}({\bf k}, t)~ e^{-ikt}
- u_{r}^{\dag}({\bf k}, t)~e^{ikt}]~\chi ({\bf x},t) ~,\\
\lab{10} i \frac{\partial u_{r}({\bf k}, t)}{\partial t} &=& i d_{e}
\sqrt{\rho} \sqrt{\frac{k}{2}}  e^{ikt} \int d \Om ({\bf \ep}_{r}
\cdot {\bf x}) |\chi ({\bf x},t)|^{2}, \eea
where $d_{e}$ is the magnitude of the electric dipole moment, $\rho
\equiv \frac{N}{V}$, and ${\bf \ep}_{r}$ is the polarization vector
of the em mode (for which the transversality condition ${\bf k}
\cdot {\bf \ep}_{r} = 0$ is assumed to hold). In Eqs. (\ref{10}) the coupling $d_{e}
\sqrt{\rho}$ is enhanced by the factor $\sqrt{N}$ due to the rescaling of
the fields. More comments on this point are included later on.

Eqs.~(\ref{17a}) - (\ref{17c}), which are derived from Eqs. (\ref{10}) and presented in Appendix C, show that the strength of the coupling between the em modes and the molecular levels is given by $\Om = {4ed_{e}}\sqrt{{N}/({6\om_{0}V})}~\om_{0} \equiv G~\om_{0}$. For pure water under standard conditions, the dimensionless quantity $G \simeq 13$. The coupling $\Om$ scales with $\sqrt{\rho}$ and thus varies with temperature and pressure. The coupled equations (\ref{17a}) - (\ref{17c}) admit the constants of motion $Q$ (see Eq. (\ref{4b})) and the quantity 
\be \lab{15} |u(t)|^{2} + 2~|a_{1}(t)|^{2} =
\frac{2}{3}\sin^{2}\theta_{0}. \ee
For further details, see Appendix C and Refs. \cite{PRL1988, PRA2006, Blasone2011}.

Consistency of Eq.~(\ref{15}) with the initial condition (\ref{4f}) implies that at $t=0$ we have
\be \lab{4f2} |u(0)|^{2} = 0. \ee
Since $|u(t)|^{2} \geq 0$, it also implies that
$|a_{1}(t)|^{2} \leq \frac{1}{3}\sin^{2}\theta_{0}$ and therefore, due to (\ref{4b}),
$|a_{0}(t)|^{2} \geq \cos^{2}\theta_{0}$.

Let us highlight a few consequences of our analysis. In our initial conditions,
the value $a_{0}(0) = 0$ has been excluded (see comments following
Eq. (\ref{4f})), on the basis of physical considerations. It is remarkable
that the dynamics {\it self-consistently} excludes such a value: $a_{0}(t) = 0$ appears to be, for any $t$, the relative maximum for the potential (see Appendix C),
and therefore, for any $t$, it is an instability point away from which the system
(spontaneously) runs. Consistently,  use of the
constants of motion (\ref{4b}) and (\ref{15}) shows that $|a_{0}(t)|^{2}$ cannot be zero, since
$|a_{0}(t)|^{2} = 0$ would imply $U^{2}(t) = - \frac{2}{3}\cos^{2}\theta_{0}$,
which is not possible since $U(t)$ is real.

Next we observe that, due to the spontaneous symmetry breaking, the system,  moving away
from the initial condition values (\ref{4f}) and (\ref{4f2}), dynamically tends to the circle of
squared radius $\ga_{0}^{2}(\theta_{0})$ (Eq. (\ref{29})),
where the nonvanishing time-independent values $A^{2}_{1} = \frac{1}{6} \sin^{2} \theta_{0}$ and
$\overline{U}^{2} = \frac{1}{3} \sin^{2} \theta_{0}$ are obtained (see Eqs. (\ref{4b}) and
(\ref{15})). The bound  $|a_{0}(t)|^{2} \geq \cos^{2}\theta_{0}$ discussed after Eq. (\ref{15}) is dynamically satisfied by $|a_{0}(t)|^{2} = \ga_{0}^{2}(\theta_{0})$ (see Eq. (\ref{29}), which actually implies $|a_{0}(t)|^{2} > \cos^{2}\theta_{0}$).

The discussion presented in Appendix C regarding the em mode $u (t)$ shows that the system runs away from the state with $u (t) = 0$ for any $t$, which is also consistent with the dynamics of the modes $a_{0}(t)$ and $a_{1}(t)$. The possibility $\theta_{0} \leq {\pi}/{4}$ is thus excluded. It is also shown in this appendix that dynamical consistency requires that  $|u(t)|^{2}  = - ({1}/{3}) \cos 2\theta_{0} \equiv v^{2}(\theta_{0})$, with  $\theta_{0} > {\pi}/{4}$, which is time-independent (see Eq.~(\ref{23})). One thus realizes that a coherent em field pattern (``limit cycle'' state) is generated as an effect of the spontaneous breakdown of the ($U(1)$) phase symmetry. This is further discussed in the next section, where we also discuss the interaction of the enzyme dipole field with the collective dipole wave field.

\section{Interaction between Enzyme and Dipole Wave Field}

In order to study how the enzyme radiative dipole field interacts with the em field pattern described above, let us first discuss a few more consequences emerging from the results we have obtained so far. We remark that Eq. (\ref{23}) implies the existence of a time-independent amplitude $\overline{U}$ for the field $u(t)$.
Inspection of the motion equations (\ref{17a})-(\ref{17c}) shows indeed that
\bea \lab{24a}  \dot{U}(t) &=& 2\Om A_{0}(t)A_{1}(t) \cos \al (t) ~\\
\lab{24b} \dot{\varphi}(t) &=& 2\Om
\frac{A_{0}(t)A_{1}(t)}{U(t)}\sin \al (t), \eea
where $\al = \de_{1}(t) - \de_{0}(t) -
\varphi(t)$.
We thus see that $\dot{U}(t) = 0$, i.e., a time-independent
amplitude $\overline{U}$ exists, if and only if the phase-locking relation
\be \lab{25} \al = \de_{1}(t) - \de_{0}(t) - \varphi (t) =
\frac{\pi}{2} \ee
is satisfied. This relation can be recast as
\be \lab{26}  \dot{\varphi}(t) =  \dot{\de_{1}}(t) -
\dot{\de_{0}}(t) = \om,\ee
with $\om$ introduced in Eq.~(\ref{8}) and proportional to $\Om$ when Eq.~(\ref{25}) holds.

In such a dynamical regime, the so-called ``limit cycle'' regime, a central result is that $A_{0}^{2} - A_{1}^{2}$ is also time-independent. Indeed, by solving the system of equations (\ref{4b}) and (\ref{15}) for $|a_{0}(t)|^{2}$ and $|a_{1}(t)|^{2}$ in terms of $|u(t)|^{2}$ and then subtracting, we obtain
\be \lab{26g} \overline{A}_{0}^{2} - \overline{A}_{1}^{2} =
\cos^{2} \theta_{0} - \frac{1}{3}\sin^{2} \theta_{0} + 2\overline{U}^{2} \neq 0, \ee
to be compared with $A_{0}^{2}(t) - A_{1}^{2}(t) \approx 0$ at thermal equilibrium in the absence of collective dynamics.

Eqs. (\ref{25}) and (\ref{26}) provide the formal expression of these collective dynamics: the system comprised of the em field interacting with the molecular water dipoles is characterized by the ``in-phase" (coherent) dynamics expressed by the phase-locking Eq. (\ref{25}). The meaning of Eq. (\ref{26}) is that any change in time of the difference between the
phases of the amplitudes $a_{1}(t)$ and $a_{0}(t)$  is compensated
by the change of phase of the em field. In other words, the gauge invariance of the theory
is preserved by the dynamical emergence of coherence between the
matter field (DNA-water-enzyme) and the em field (radiative dipole field from DNA, water, and enzyme).

We note here an estimate of the resulting interaction energy. For $ 0 < \theta_0 < \pi/3$, the average pulsations (see Appendix C, especially Eqs. (\ref{20}) and (\ref{23})) fall in the range $0 < \nu < 1000 \, \text{cm}^{-1}$, populating bands in the infrared spectrum \cite{PRL1988} and overlapping with the lower end of the 0.1 - 1 eV range mentioned in Section II. This is a remarkable result, considering the rough analytical form of our water model, without an abundance of external parameters. These wavelengths, which overlap with the energy scale of collective dipole modes in DNA and enzyme systems, provide an estimate of the long-range interactions mediated by the collective water dipole field, and they are distinct from the more energetic absorption bands of water due to intramolecular vibrations and electronic transitions. These long-wavelength interactions do not exist in pure water, as evidenced by Eqs. (\ref{26g}) and (\ref{4pb}). Water adapts itself to biological agents (e.g., DNA and enzyme) and compensates, through gauge invariance, by phase-locking its electromagnetic field---arising from water's matter field of endowed electric dipoles---to the field introduced by the biological agents.

In summary, as a consequence of the interaction with DNA and the water's own radiative em field, the combined system evolves away from the initial symmetric vacuum to the asymmetric vacuum $|u(t)|^{2}  \neq 0$. The coherent em field pattern that emerges is formally expressed by the phase locking in Eqs. (\ref{25}) and (\ref{26}).

We consider now the polarization density of the water environment in the presence of the enzyme radiative electric field ${\bf E}$, assumed parallel to the ${\bf z}$ axis, as chosen in Section III. The term
\begin{equation} 
\label{4p} {\cal H} = - {\bf d_{e}} \cdot {\bf E}, 
\end{equation}
with ${\bf d_{e}}$ the electric dipole moment for water, is then added to the system energy.

By following Ref. \cite{Knight}, it is possible to  write the interaction Hamiltonian  ${\cal H}$ of Eq.~(\ref{4p}) in the form of a Jaynes-Cummings-like Hamiltonian for large $N$:
\be \lab{JC} {\cal H} = \hbar \sqrt{N}\ga (a^{\dag} S^{-} + ~a S^{+}),
\ee
where $\ga$ is the coupling constant proportional to the matrix
element of the molecular dipole moment  and to the inverse of the
volume square root $V^{-1/2}$, $a^\dagger$ and $a$ are the creation and annihilation operators discussed in Figure (\ref{enzcohmodes}) for the electric field ${\bf E}$, and $S^{\pm}$ denote the creation and annihilation operators of the ``dipole wave modes'' of the field pattern imaging the DNA-water interaction.

Note that the factor $\sqrt{N}$ multiplying the coupling $\ga$ implies that for large $N$ the collective mesoscopic interaction energy scale is much larger, by the factor $\sqrt{N}$, than the microscopic fluctuations due to individual water molecules, thus providing a protective gap against thermalization for the long-range quantum correlations. This fact is well known in studies of superradiance and cooperative robustness for aggregates of two-level systems \cite{Celardo1, Celardo2}. Conversely, the time scale is much shorter, by the factor $1/\sqrt{N}$, than
the short-range interactions among the molecules. A similar remark applies to the enhancement of the coupling $G$, also due to the multiplying factor $\sqrt{N}$ (see the comments after Eq.~(\ref{4f})). In this vein, we observe that in our discussion we have not considered the losses of the radiative energy
from the volume $V$. One can estimate these losses by comparing lifetimes of the different modes, namely considering the different time scales associated with them, ${2\pi}/\om_{0}$, ${2\pi}/\om_{1}$, ${2\pi}/{\overline{\om}}$, and
${2\pi}/\om$, where we have put $\om_{1} \equiv m  = 2\sqrt{2}{\Om} \sqrt[4]{1 - (1/4)\sin^{2} 2\theta_{0}}$, in the limit cycle regime $\theta_0 > \pi/4$, and  ${\overline{\om}} = \Om \sqrt{2 \cos 2\theta_{0}}$ for $0 < \theta_0 <  \pi/4$ (see Appendix C). Thus again we see that for large $N$ the
time scale of the collective interaction is much shorter by the factor $1/\sqrt{N}$
than the short-range interactions among the water molecules. It then follows that the macroscopic stability of the system is protected against quantum fluctuations in the microscopic short-range dynamics. In a similar way, for sufficiently
large $N$ the collective interaction is protected against thermal
fluctuations, which may affect the
collective process only when $k_{B} T$ ($\sim$.02 eV at physiological temperatures) is of the same order or larger than the energy gap determining the height of the protection.

The interaction ${\cal H}$ produces mixing between the states
$Y^{0}_{0}$ and $Y^{0}_{1}$ (see Eq.~(\ref{4a})): $Y^{0}_{0} \rightarrow Y^{0}_{0} \cos
\tau + Y^{0}_{1} \sin \tau$ and $Y^{0}_{1} \rightarrow - Y^{0}_{0}
\sin \tau + Y^{0}_{1} \cos \tau$, with
\be \lab{4pa} \tan \tau = \frac{\om_{0} - \sqrt{\om_{0}^{2} + 4
{\cal H}^{2}}}{2{\cal H}}. \end{equation}
The polarization $P_{{\bf n}}$ is now found to be
\bea \nonumber P_{{\bf n}} &=& \frac{1}{\sqrt{3}} (\overline{A}_{0}^{2} - \overline{A}_{1}^{2})
\sin 2\tau \\
 &+&  \frac{2}{\sqrt{3}} \overline{A}_{0}^{2}\overline{A}_{1}^{2} \cos 2\tau  \cos [(\om - \om_{\cal H})t],  \lab{4pb}  \eea
where $\om_{\cal H} \equiv \sqrt{\om_{0}^{2} + 4 {\cal H}^{2}}$ and the limit cycle amplitudes $\overline{A}_{0}^{2}$ and $\overline{A}_{1}^{2}$ have been used.  Time averaging gives the polarization $\overline{P_{{\bf n}}} = (1/\sqrt{3}) (\overline{A}_{0}^{2} - \overline{A}_{1}^{2}) \sin 2\tau$.  $\overline{P_{{\bf n}}}$ is non-zero as long as $\tau \neq 0$ and provided that the difference  $(\overline{A}_{0}^{2} - \overline{A}_{1}^{2})$ is non-vanishing.
As we have seen, the former condition ($\tau \neq 0$) is realized due to the presence of the enzyme electric dipole field, and the latter condition ($\overline{A}_{0}^{2} - \overline{A}_{1}^{2} \neq 0$) is realized in the water coherent domains imaging the DNA em radiative field. $\overline{P_{{\bf n}}}$ thus carries information on both the DNA radiative field and the enzyme field, and it is vanishing if one or both fields are absent. 

We would like to clarify how frequencies in the 0.1 - 1 eV range can effectively couple to rotational transitions of liquid water. Though the rotational transitions of individual water molecules correspond to energies orders of magnitude smaller (meV), the point we would like to emphasize is that the dipole-dipole modes of the aromatic networks do not couple to the rotational transition energies of individual water molecules but rather to the collective polarization modes present in the molecular water dipole field. The dipole-dipole modes of the aromatic networks become sources for radiative fields that stimulate this collective polarization, and thus, because this is a field effect, the electronic polarizability of individual water molecules is not involved. See, in particular, Eq. (\ref{8}), which shows the collective polarization as a function of the phase shift $(\omega - \omega_0)t$ between the water dipole field and an individual water molecule. Please note that these polarization modes have been studied in the formalism of quantum electrodynamics since at least the 1980s \cite{PRL1988}. 

Furthermore, the infrared spectrum of liquid water (i.e., the 0.1 - 1 eV range) is dominated by the intense absorption due to fundamental O-H stretching vibrations. Though there is no rotational fine structure in this region, the absorption bands are broader than might be expected because of collective behaviors in the water dipole field. Peak maxima for liquid water are observed at 3450 cm$^{-1}$ (0.43 eV), 3615 cm$^{-1}$ (0.45 eV), and 1640 cm$^{-1}$ (0.20 eV), completely within the 0.1 - 1 eV range. Liquid water also has absorption bands around 5128 cm$^{-1}$ (0.64 eV), 6896 cm$^{-1}$ (0.86 eV), 8333 cm$^{-1}$ (1.03 eV), and 10300 cm$^{-1}$ (1.28 eV). 

The collective polarization modes we describe are Nambu-Goldstone (NG) modes, which are generated in spontaneous symmetry breaking processes. Specifically, in our case, there is a breakdown of rotational dipole symmetry. In the infinite volume limit, NG modes are massless; however, as an effect of the system boundaries and defects, NG modes acquire a non-zero mass. The NG modes with nearly vanishing mass, as dictated by the Goldstone theorem, are low-energy (low-momentum) modes and thus couple to the longer wavelength range of the spectrum mentioned above. Indeed, in their condensation in the lowest energy state (the ground state), nearly vanishing momentum $k$ implies due to the De Broglie relation $(k = h/\lambda)$ that low-energy NG modes are quanta associated with long wavelengths ($\sim$10-100 microns, much greater than the size of an individual water molecule). Namely, they contribute to the infrared transitions of the system mentioned above. 

The popular metaphorical expression that ``the enzyme sees the DNA'' thus acquires the exact physical meaning that the enzyme couples to the collective dipole wave mode dynamically generated by the radiative em interaction between DNA and its water environment. What the enzyme ``sees'' is the em image of DNA in the water environment. We thus obtain a water-mediated picture of DNA-enzyme interactions. The underlying structure of such a picture is the standard one of the QFT interaction between two systems: they ``see'' each other by exchanging a mediating correlation wave or quantum (e.g., the photon in quantum electrodynamics). In the present case,  the dipole wave of the water polarization pattern plays the role of the mediating correlation mode.

We stress that our analysis does not undermine traditional biochemical analysis, which remains valid in its purely statistical description of average regularities in living systems. What our analysis adds to DNA biochemistry is a careful consideration of the mediating radiation field in which DNA biochemical processes occur. Traditional stoichiometric and thermodynamics-based analysis remains valid within the relevant domains of applicability. What our results provide is a deeper understanding of the dynamical processes orchestrating long-range em correlation modes in biology, which can account for the remarkable efficiency, space-time ordering, and diverse time scales of these otherwise fully stochastic molecular activities.

\section{Conclusions}

We have considered the interaction between DNA and enzyme macromolecules in their water environment. By resorting to recent results \cite{Kurian}, we have discussed the collective dipole behavior and the Hamiltonians of DNA and enzyme molecules. We have considered the \textit{Eco}RI restriction endonuclease and the \textit{Taq} DNA polymerase, which are used widely in molecular biology for precise cutting and rapid amplification of DNA sequences, respectively. We have adopted the QFT paradigm, according to which any interaction between two systems is mediated by the propagation of a mediating correlation field or quantum, such as, for example, the photon exchanged between interacting electric charges in quantum electrodynamics. In the present study of DNA-enzyme interactions, we have identified such a correlation field with the collective dipole wave of the molecular water field. The DNA radiative dipole field does indeed trigger the collective dipole wave in the water environment, which in turn couples with the enzyme radiative dipole field. By following Ref. \cite{PRA2006}, we have shown that a nonzero time-independent em amplitude can develop, driving a phase transition, as an effect of the radiative dipole-dipole interaction. We have thus discussed the transition of the molecular water field to the limit cycle regime, where the em DNA dipole field and the water dipole field get locked in phase. We have shown that the specific value of the resulting em dipole field amplitude controls the boundary conditions of the dynamics and the characteristic time scales of the process. These time scales turn out to be much shorter than the ones associated with thermal noise, providing a protective energy gap for the collective dynamics.

While a comparison with the strength of other interactions involved---electrostatic, hydrophobic, etc.---is possible, it is not relevant to our study because these interactions occur at short range. Water has a unique ability to shield charged species from each other, so electrostatic interactions between charges are highly attenuated in water. The electrostatic force between two charges in solution is inversely proportional to the dielectric constant of the solvent. The dielectric constant of water (80.0) is huge, over twice that of methanol (33.1) and over five times that of ammonia (15.5). Water is therefore a good solvent for salts because the attractive forces between cations and anions are minimized. In this respect, it is interesting to observe that the decoherence time for ion-ion collisions and for interactions with distant ions in aqueous solutions for crystalline binary compounds is of the order of $10^{-40}$ and $10^{-38}$ of a second, respectively \cite{DecoherenceCriterion2001}. Such very short decoherence times would make impossible the formation of salt crystals, which of course contradicts common experience, showing that in practice the formation of stable salt crystals does occur and lasts for many orders of magnitude longer than $10^{-40}$ and $10^{-38}$ of a second. At the crystallization point and low temperature, this formation can occur within fractions of a second to several seconds or even longer times, from minutes to hours. 

The way out of the contradiction lies in the fact that the binding of the atoms in the crystalline lattice is not the result of ion-ion collisions or of interactions with distant ions in a quantum mechanical scheme. The atom binding is instead due to long-range correlations mediated by Nambu-Goldstone modes in a quantum electrodynamics framework. In the case of crystal lattices, these modes are indeed the phonons, and they are known to be extremely important in many areas of solid state physics~\cite{Feynman,Umezawa}. 

It has also long been recognized that hydrogen bonding is a dominant mechanism of cohesion in water systems, and some of the failures of previous models have originated in the exclusion of crucially important dispersion interactions. However, hydrogen bonding is a complex phenomenon, which may be decomposed into electrostatic attraction, polarization, dispersion, and partial covalency, etc., even though the relative contributions of these components in water remain controversial, depending significantly on definition. The contribution due to partial covalency (often described as charge transfer), for example, has been particularly contentious. Hydrophobic effects, furthermore, are a consequence of strong directional interactions between water molecules and are driven by entropic considerations in bulk water. In contrast, the phenomenon we describe in this manuscript is a mesoscopic effect that occurs due to the long-range radiative correlations of water at biopolymeric interfaces. Several estimates place the strengths of these various contributions due to water dimer interactions at the meV scale \cite{DFT4H20}, various orders of magnitude smaller than the collective dipole modes described in our manuscript, and therefore too small to couple effectively. Our contention, based on estimates in this paper, is that collective dipole modes in the 0.1 - 1 eV range may serve a critical role in facilitating interactions between DNA and (endonuclease or polymerase) enzyme systems. 

We observe that the system ground state is a coherent condensate of the massless modes (of Nambu-Goldstone type), which have been identified in our analysis in Appendix C. We have indeed seen that the cylindrical ($U(1)$) phase symmetry gets spontaneously broken. As a final consequence, then, the phase-locking relation in Eqs. (\ref{25}) and (\ref{26}) is obtained. The consistency of such a complex dynamical process in biology with the gauge invariance of QFT can be understood as follows: Gauge invariance requires fixing the phase of the matter field and implies that a specific gauge function must be selected. This leads to the introduction of the covariant derivative $D_{\mu} = \pa_{\mu} - igA_{\mu}$. The variations in the phase of the matter field thus give origin to the pure gauge $A_{\mu} = \pa_{\mu} \varphi$. This means that when $\varphi({\bf x},t)$ is a regular (continuous and differentiable) function, then ${\bf E}=-\frac{\pa {\bf A}}{\pa t} +
{\bf \nabla} A_{0}= (-\frac{\pa }{\pa t}{\bf \nabla} +{\bf \nabla}
\frac{\pa }{\pa t})\varphi  = 0$, since time derivative and gradient operator can be interchanged for regular functions. In such a case, we also have ${\bf B}={\bf \nabla} \times {\bf A} = {\bf \nabla}
\times {\bf \nabla} \varphi= 0$. Therefore, the fields {\bf E} and  {\bf B}  can acquire nonvanishing values
in a coherent region only if $\varphi({\bf x},t)$
exhibits a divergence or a topological singularity within the
region \cite{Alfinito:2001mm}. Such an occurrence is particularly relevant in our analysis, since it accounts for topologically nontrivial configurations in the coherent ground state, thus requiring the ``clamping'' characterizing a wide swath of restriction endonuclease and DNA polymerase catalytic activity. \textit{Eco}RI locates its target DNA sequence of six base pairs via ``facilitated diffusion'' in a non-specific conformation that is characterized by interstitial water molecules between the DNA and enzyme. In this conformation and under optimum conditions, these enzymes are able to scan up to $10^6$ base pairs in a single binding event. Upon recognizing its target sequence, \textit{Eco}RI changes to a specific conformation that tightly binds the DNA through exclusion of the interstitial water. In the $Taq$ DNA clamp, a protein multimeric structure completely encircles the DNA double helix, essentially encircling the core of a vortex, which is formed not coincidentally by a layer of water molecules in the central pore of the clamp between the DNA and the protein surface. As is well known \cite{NuclPhys75, NuclPhys1975}, the dynamical formation of vortex structures is described by the QFT of spontaneously broken $U(1)$ symmetry theories.

These findings may contribute toward addressing an unsolved problem in biology: Why are divalent cations like Mg$^{2+}$ so important to enzyme behavior and so precisely controlled for optimum biological functions? For example, \textit{Eco}RV (a close relative of \textit{Eco}RI) incubated at ideal pH cuts both strands of DNA in the synchronized, concerted manner discussed in Ref \cite{Kurian}; in contrast, the enzyme reaction at lower pH involves sequential, independent cutting of the two strands \cite{halford1988modes}. The difference in catalysis, which cannot be accounted for by weakened enzyme-DNA binding, has been traced to the asymmetrical binding of Mg$^{2+}$ to the \textit{Eco}RV subunits. Thus we see that the pH disturbance propagates through the buffer solution, generating a local electromagnetic environment in which the symmetry of the complex is broken. Similarly, \textit{Taq} is extremely dependent on magnesium, and determining the optimum concentration to use is critical to the success of the polymerase chain reaction. Just as complex systems exhibit behavior that cannot be predicted from the mechanics of microscopic constituents, so biology has dynamically optimized several parameters to achieve maintenance of long-range correlations. Divalent cations may therefore serve as electromagnetic antennae to enhance or maintain mediating dipole-wave fields in solution.

In conclusion, both the high efficiency and high reliability of enzymatic catalytic activity in DNA metabolism rest on the physico-chemical properties of enzymatic macromolecules, which aid in identifying the water em field pattern imaging the DNA molecules in their water environment.

\begin{acknowledgments}
AC and GV acknowledge partial financial support from MIUR and INFN. TJAC would like to acknowledge financial support from the Department of Psychology and Neuroscience and the Institute for Neuro-Immune Medicine at Nova Southeastern University (NSU), and work in conjunction with the NSU President's Faculty Research and Development Grant (PFRDG) program PFRDG 335426 (Craddock -- PI). PK was supported in part by the National Center For Advancing Translational Sciences of the National Institutes of Health under Award Number TL1TR001431. The content of research reported in this publication is solely the responsibility of the authors and does not necessarily represent the official views of the National Institutes of Health. PK would also like to acknowledge ongoing discussions with G. Dunston, as well as partial financial support from the Whole Genome Science Foundation. 
\end{acknowledgments}

\appendix
\section{}

The revolution in genetic sequencing and engineering has been made possible through the use of restriction endonucleases. Originally isolated from bacteria, these enzymes cut DNA at recognition sequences with high specificity, thereby assuring a consistent final product. Each endonuclease has been named using a system loosely based on the bacterial genus, species, and strain from which the enzyme is derived: the first identified endonuclease in the RY13 strain of \textit{Escherichia coli} is \textit{Eco}RI, and the fifth endonuclease extracted from the same strain is \textit{Eco}RV. The body of literature on the structure and catalytic mechanisms of these molecular workhorses is substantial and has been reviewed in multiple instances \cite{modrich1982studies, jeltsch1995evidence, jen1998protein, pingoud2001structure, pingoud2002evolutionary, roberts2003nomenclature, pingoud2005specificity, pingoud2005type}.

One study \cite{stahl1996introduction} revealed that the \textit{Eco}RV DNA-binding domains cannot function independently of each other, and that only with asymmetric modifications can an \textit{Eco}RV mutant cleave DNA in a single strand of the recognition site. Also, \textit{Eco}RV mutants are not affected in ground state binding but rather in the stabilization of the transition state, and catalysis is significantly altered compared to binding when the symmetry of the protein-DNA interface is disturbed. Taken together, these data suggest that an asymmetry in the enzyme is manifested in the catalytic centers of the two subunits only in the transition state, and that a nonlocal pathway---in which some physical quantity is conserved---may be used for coordination.

The notion that enzyme catalysis can be ``substrate-assisted'' is not new \cite{jeltsch1995evidence, horton1998metal}, and previous authors have hypothesized that energy could be transferred from enzyme clamping to the catalytic transition state \cite{jen1998protein}, though no quantitative mechanisms were proposed. The idea that enzymes may sequester coherent energy from DNA zero-point modes for genomic metabolism \cite{Kurian} certainly fills a gap in our understanding of how enzyme clamping on DNA might give rise to energy transport in the substrate, which then ``assists'' in synchronization of the catalytic process to form double-strand breaks.

This process of substrate-assisted catalysis occurs after the \textit{Eco}RI enzyme has bound to its cognate DNA sequence and changed conformation so as to be in the catalytic transition state. Thus, already having overcome the activation energy barrier, the reaction proceeds spontaneously due to the negative change in Gibbs free energy, which is provided here as an estimate for the hydrolysis of a phosphodiester bond ($\varepsilon_{P-O}$). The collective dipole-dipole oscillation mediates the synchronized catalysis of two phosphodiester bonds at about 20 \r{A} spatial separation and ensures that the spontaneous Gibbs free energy change is channeled to the appropriate catalytic center in a symmetric way. Since the \textit{Eco}RI enzyme does not use ATP for its catalytic activity, this energy is recruited from the collective oscillation after the enzyme has already overcome the activation energy barrier due to conformation change upon target sequence recognition.

Developed in 1983 by Kary Mullis and colleagues \cite{Mullis}, the polymerase chain reaction (PCR) is now a common and often indispensable technique used in biological and medical research, where a single copy or few copies of DNA must be amplified by several orders of magnitude. The method relies on thermal cycling, which consists of repeated stages of 1) DNA denaturation (into single strands), 2) primer annealing, and 3) DNA replication. This last stage is guided by a heat-tolerant DNA polymerase called \textit{Taq}, which was originally isolated from the thermophilic bacterium \textit{Thermus aquaticus}.  \textit{Taq} functions optimally between $75$ and $80^\circ$C and enzymatically assembles a new DNA strand from nucleotides that are added to the PCR solution. Specially designed primers complementary to the target region ensure amplification of only the desired DNA sequence under consideration. As the target sequence is doubled during each cycle, available substrates in the reaction can become limiting as the PCR progresses, usually between 20 and 40 cycles.

\section{}

Using matrix elements for the derivation of the base-pair electronic angular frequencies, by using ${(\gv{\pi})}_{t} = Qr_t$ and $\mathbf{(F)}_{u} = {k_{uv}r_v}$ we write $\gv{\pi} =\gv{\alpha}\cdot \mathbf{E} = \gv{\alpha}\cdot {\mathbf{F}}/{Q}$ as $Qr_t = \alpha_{tu}{k_{uv}r_v}/{Q}$, or $Q^2 \delta_{tv} = \alpha_{tu}k_{uv}$, and thus
\begin{eqnarray}
Q^2\gv{\alpha}^{-1} =\mathbf{k} = m\gv{\omega}^2,
\end{eqnarray}
which yields precisely the matrix version of Eq.~(3) in the main text.

In the case of an infinite helix composed of homogeneous base pairs, we may transform the infinite sums into easily calculable integrals. From the idealized Hamiltonian
\begin{align}
H_I= \frac{1}{2M} \left \{ \sum \limits_{n=-\infty}^{+\infty} p_n^2 \right \} &+ \frac{1}{2}M \omega^2 \left \{ \sum \limits_{n=-\infty}^{+\infty} q_n^2 \right \} \\ \nonumber
&+\frac{1}{2}\Gamma \left \{ \sum \limits_{n=-\infty}^{+\infty} (q_n-q_{n+1})^2 \right \},
\end{align}
where the uniform $M$ and $\omega$ reflect the chain homogeneity and $p_n(t), q_n(t)$ are the deviation from equilibrium for the $n$th component oscillator in a single dimension with arbitrary interaction potential $\Gamma$, we may use the Bessel-Parseval relation \cite{cohen1977quantum} and recursion to obtain the diagonalized form for $H_I$:
\begin{align} \label{eq:normalmodes}
\int \limits_{-\frac{\pi}{\ell}}^{+\frac{\pi}{\ell}}\frac{\ell  dk}{2\pi}  \left\{\frac{|P_0(k, t)|^2}{2M} + \frac{M}{2} \left [\omega^2 + \frac{4\Gamma}{M} \sin^2\left (\frac{k \ell}{2} \right) \right] |Q_0(k, t)|^2 \right\}
\end{align}
where $\ell$ is the unit distance of the oscillator chain and the new position and momenta coordinates are defined using Fourier series:
\begin{eqnarray} \label{newp&q}
\begin{gathered}
Q_0 = \sum \limits_{n=-\infty}^{+\infty} q_n e^{-ink \ell},\\
P_0 = \sum \limits_{n=-\infty}^{+\infty} p_n e^{-ink \ell},\\
\end{gathered}
\hspace{0.25em}
\begin{gathered}
Q_1 = \sum \limits_{n=-\infty}^{+\infty} q_{n+1} e^{-ink \ell},...\\
P_1 = \sum \limits_{n=-\infty}^{+\infty} p_{n+1} e^{-ink \ell},...\\
\end{gathered}.
\end{eqnarray}

\begin{figure*}
	\begin{subfigure}[b]{0.5\linewidth}
          	\includegraphics[width=\textwidth]{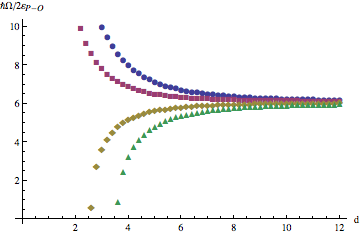}
		\subcaption{dsAAAA longitudinal modes} \label{fig:AAAAz}
        \end{subfigure}%
        \begin{subfigure}[b]{0.5\linewidth}
        		\includegraphics[width=\textwidth]{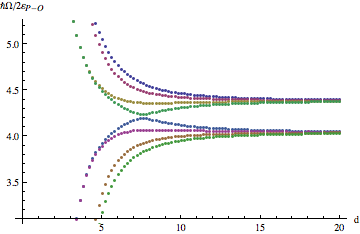}
          	\subcaption{dsAAAA transverse modes} \label{fig:AAAAxy}
        \end{subfigure}   
\captionsetup{justification=raggedright, singlelinecheck=false}
\caption{\textbf{Zero-point modes of DNA sequence.} Longitudinal (\ref{fig:AAAAz}) and transverse (\ref{fig:AAAAxy}) zero-point modes parametrized by the inter-base-pair spacing $d$ for the double-stranded DNA sequence AAAA. The abscissae are given in \r{A} and the ordinates are dimensionless. Notice the divergence of the zero-point modes at the observed equilibrium base-pair spacing $d=3.4\,$\r{A}. In six- and eight-bp palindromic sequences, this structural requirement on the architecture of the double helix ``tunes" DNA collective dipole oscillations to the resonant energy required for synchronized double-strand breakage, $2\varepsilon_{P-O} \simeq 0.46$ eV. DNA sequence is therefore exquisitely structured to channel its own electronic vibrational energy for the emergence and continuity of biological diversity and other life processes.}
\label{subfig:dsAAAA}
\end{figure*}

The normal modes of vibration are obtained readily from Eq.~(\ref{eq:normalmodes}): $\Omega(k)=\sqrt{\omega^2 + \frac{4\Gamma}{M} \sin^2\left ( \frac{k\ell}{2} \right)}$, which applies for any infinite coupled oscillator sequence with homogeneous components in the first Brillouin zone, where the wave vector obeys $-\frac{\pi}{\ell} \leq k \leq \frac{\pi}{\ell}$. Transforming coordinates by Fourier expansion thus allows us to convert the idealized Hamiltonian, which was expressed as an infinite sum of coupled oscillators, into a definite integral over uncoupled collective modes for each physically distinguishable state.

Though a helpful comparison, such an analytical solution is not rigorously applicable for finite, real DNA sequences. It does not show the complement of modes depressed below the homogeneous trapping frequency $\omega$ because the generality of the real-valued function $\Omega$ would have to be restricted for specified values of $\Gamma$. However, numerical analysis will certainly suffice for genomic length scales. 

With reference to our computations for DNA sequences discussed in Section II, we note that by separating Eq.~(\ref{eq:Hamiltonian}) into energy contributions from transverse ($H_{xy}$) and longitudinal ($H_z$) modes, we may write the symmetric longitudinal potential matrix $\mathbf{V}_z$ for a four-bp sequence as
\begin{equation}
{\begin{pmatrix}
k_{1,zz} & \gamma_{12}^z & 0 & 0 \\
\gamma_{21}^z  & k_{2,zz} & \gamma_{23}^z  & 0 \\
0 & \gamma_{32}^z  & k_{3,zz} & \gamma_{34}^z  \\
0 & 0 & \gamma_{43}^z  & k_{4,zz} \\
\end{pmatrix}},
\end{equation}
where $k_{1,zz}=m_1\omega_{1,zz}^2$, etc., and $\gamma_{s,s+1}^z=\gamma_{s+1,s}^z=-Q^2/(2\pi\epsilon_0 d^3)$ denotes the $z_s z_{s+1}$ coefficient from Eq.~(\ref{eq:dipolepotential}). The symmetric transverse potential matrix $\mathbf{V}_{xy}$ is
\begin{equation}
{\begin{pmatrix}
k_{1,xx}& \gamma_{12}^x & 0 & 0 &0 & \gamma_{12}^{xy} &0 &0\\
\gamma_{21}^x  & k_{2,xx} & \gamma_{23}^x  & 0  & \gamma_{21}^{xy} &0 &\gamma_{23}^{xy}  &0\\
0 & \gamma_{32}^x  & k_{3,xx} & \gamma_{34}^x &0 &  \gamma_{32}^{xy}  &0 & \gamma_{34}^{xy}\\
0 & 0 & \gamma_{43}^x  & k_{4,xx} &0 & 0& \gamma_{43}^{xy} & 0 \\
0&\gamma_{12}^{yx}&0&0&k_{1,yy}&\gamma_{12}^y&0&0\\
\gamma_{21}^{yx}&0&\gamma_{23}^{yx}&0&\gamma_{21}^y&k_{2,yy}&\gamma_{23}^y&0\\
0&\gamma_{32}^{yx} &0&\gamma_{34}^{yx} &0&\gamma_{32}^y&k_{3,yy}&\gamma_{34}^y\\
0&0&\gamma_{43}^{yx} &0&0&0&\gamma_{43}^y&k_{4,yy}\\
\end{pmatrix}},
\end{equation}
where $k_{1,xx}=m_1\omega_{1,xx}^2$, etc., and $\gamma_{12}^{xy} =-Q^2 \sin\theta/(4\pi\epsilon_0 d^3)=-\gamma_{21}^{xy}$, etc. The diagonal kinetic matrices $\mathbf{T}_j$ consist of the electronic oscillator masses:
\begin{align}
\mathbf{T}_z &= \text{diag}(m_1, m_2, m_3, m_4), \\ \nonumber
\mathbf{T}_{xy} &=
\text{diag}(m_1, m_2, m_3, m_4, m_1, m_2, m_3, m_4).
\end{align}

Parametrization of our model by the inter-base-pair spacing is shown in Figure \ref{subfig:dsAAAA} for a homogeneous dsDNA sequence of four base pairs. What becomes quickly apparent is the rapid convergence of the longitudinal modes to $\hbar\Omega/2 \approx 6.0 \varepsilon_{P-O}$, within less than three times the standard observed inter-base-pair spacing, suggesting that this parameter has been evolutionarily optimized to maximize variation in DNA electronic oscillations at physiologically relevant length scales. Similar behavior is observed for the transverse modes, with a bi-modal convergence around $\hbar\Omega/2 \approx 4.4 \varepsilon_{P-O}$ and $4.0 \varepsilon_{P-O}$. 

Surprisingly, the modes for AAAA converge to the same values as the triplet codon case (dsAAA, data not shown) for the zero-point energies over comparable length scales. The middle harmonics bifurcate quickly and diverge when the spacing dips below $5.0$\r{A} (longitudinal) and $7.0$\r{A} (transverse).
What these data suggest is that, if DNA were ``tuned'' away from this equilibrium spacing, then we would not observe the plethora of distinct collective vibrations in DNA nor would certain collective modes achieve energies of significance to genomic and biological metabolism. This fine-tuning of DNA architecture for coherent energy transport leads us to postulate that DNA is constructed with a view toward finding energy in its own vibrations for life processes.

We have chosen to examine the so-called ``zero-point'' modes (on which the number operator $N_{s,j}=a^{\dagger}_{s,j} a_{s,j}$ in Eq.~(\ref{eq:numop}) acts to produce a zero eigenvalue) because these are most easily excited by the free energy changes due to enzyme clamping. These zero-point oscillations are collective normal modes of the DNA system considered in our model framework. Because the oscillations are normal modes, a four-base-pair sequence will produce four frequencies of coherent (phase-synchronized) oscillation in the longitudinal direction, as shown in Eq.~(\ref{eq:goldenrat}). Similarly, a six-(eight-)base-pair sequence will produce six (eight) frequencies of coherent oscillation in the longitudinal direction, and so on. The number of frequencies of coherent oscillation in the transverse direction is doubled because of the coupling between the \textit{x} and \textit{y} degrees of freedom due to the helix twist angle.

\section{}

Taking advantage of the rotational symmetry, by using Eq.~(\ref{4a}) in Eqs.~(\ref{10}) and writing $u(t) \equiv u_{m}(t)$  and $a_{1}(t) \equiv a_{1,m}(t)$,  we get the set of equations \cite{PRA2006, PRL1988}:
\bea \lab{17a} \dot{a}_{0}(t) &=& 3~\Om~u^{*}(t)~a_{1}(t)\\
\lab{17b} \dot{a}_{1}(t) &=& - \Om~
u(t)~a_{0}(t) \\
\lab{17c} \dot{u}(t) &=&  2~\Om~ a_{0}^{*}(t)~a_{1}(t) ~, \eea
with $\Om = {4ed_{e}}\sqrt{{N}/({6\om_{0}V})}~\om_{0} \equiv G~\om_{0}$. 
These equations are fully consistent with the dipole rotational
invariance expressed by the zero average polarization, 
Eq.~(\ref{8}), the normalization condition (\ref{4}) (or (\ref{4b})), and the conservation of molecules in Eq.~(\ref{4c}). The rate of change of the amplitude of the level $(0,0)$ is shown in Eq.~(\ref{17a}) to depend on the coupling between
the levels $(1,m)$, $m = 0, \pm 1,$ and the radiative em mode of
corresponding polarization. Due to rotational invariance, each of these couplings contribute in equal measure to the transitions to
(0,0). In a similar way, the rate of change of the
amplitude of each level $(1,m)$ is shown in Eq.~(\ref{17b}) to depend on the coupling between the the level $(0,0)$ and the corresponding radiative em mode. The rate of change of
the amplitude of the radiative em mode of corresponding
polarization is controlled by the transitions $(0,0)
\leftrightarrow (1,m), m = 0, \pm 1$, as shown in Eq.~(\ref{17c}).
We remark that, as described by Eq.~(\ref{17a}),  each of the levels $(1,m)$ may find in the em field, which includes all possible polarizations, the proper mode to couple with, in full respect of the selection rules. 
Note that use of the complex conjugate of Eq.~(\ref{17b}) in
(\ref{17c}) leads to 
\be \lab{14} \frac{\pa}{\pa t} {|u(t)|^{2}} = -
2~\frac{\pa}{\pa t} {|a_{1}(t)|^{2}}.\ee
By integrating this equation and fixing the integration constants consistently with the initial conditions, Eq.~(\ref{15}) is obtained.

We  now  study in the mean field approximation \cite{ItkZuber} the ground state (vacuum) of the system for each of the modes $a_{0}(t)$, $a_{1}(t)$, and $u(t)$.
For the mode $a_{0}(t)$ we find the equation \cite{PRA2006}
\be \lab{28}  %\lab{21}
\ddot{a}_{0} (t)  = - \frac{\de }{\de
a_{0}^{*}}V[a_{0}(t), a_{0}^{*}(t)],\ee
where the potential is
\be \lab{28b} V[a_{0}(t), a_{0}^{*}(t)] = 2 {\Om}^{2}
(|a_{0}(t)|^{2} - \ga_{0}^{2}(\theta_{0}))^{2}
 ~\ee
and $\ga_{0}^{2}(\theta_{0}) \equiv \frac{1}{2}(1 + \cos^{2}
\theta_{0})$.

Denote by $a_{0,R}(t)$ and $a_{0,I}(t)$ the real
and the imaginary component, respectively, of the $a_{0}(t)$
field. The potential has a relative maximum at $a_{0}(t) =
0$ and a (continuum) set of minima on the circle of squared
radius $\ga_{0}^{2}(\theta_{0})$ in the $(a_{0,R}(t),a_{0,I}(t))$
plane, given for any t by
\be \lab{29}  |a_{0}(t)|^{2} =  \frac{1}{2} (1 + \cos^{2}
\theta_{0}) = \ga_{0}^{2}(\theta_{0}) ~. \ee
The points on the circle transform into each other under shifts of
the field $\de_{0}$: $\de_{0} \rightarrow \de_{0} + \al$
(rotations in the $(a_{0,R}(t),a_{0,I}(t))$ plane). They represent (infinitely many)
possible vacua for the system. The phase
symmetry is spontaneously broken when a specific ground state is singled out
by fixing the value of the $\de_{0}$ field. $\ga_{0}(\theta_{0})$ is the order parameter.
One can recognize that there is a quasi-periodic mode with pulsation $m  = 2{\Om} \sqrt{(1 + \cos^{2} \theta_{0})}$ and that the field $\de_{0}(t)$ corresponds to a massless mode (the so-called Nambu-Goldstone (NG) field). It is a collective mode implied by the spontaneous breakdown
of symmetry.

The motion equation for the the amplitude $a_{1}(t)$ is found to be
\be \lab{31a}  \ddot{a}_{1} (t)  =  - \frac{\de }{\de
a_{1}^{*}}V[a_{1}(t), a_{1}^{*}(t)]~,\ee
where  the
potential is
\be \lab{32}   V[a_{1}(t), a_{1}^{*}(t)] = \si^{2}|a_{1}(t)|^{2} -
6 {\Om}^{2} (|a_{1}(t)|^{2})^{2} ~\ee
and $\si^{2} = 2~{\Om}^{2}(1 + \sin^{2} \theta_{0})$.
There is a relative minimum at $a_{1} = 0$ and a
(continuum) set of  maxima on the circle of squared radius
\be \lab{32}  |a_{1}(t)|^{2}  = \frac{1}{6} (1 + \sin^{2}
\theta_{0}) \equiv \ga_{1}^{2}(\theta_{0}). \ee
For consistency between Eqs. (\ref{28}) and
(\ref{31a}), it is excluded that the  amplitude
$A_{1}$  be  zero (at the minimum of $V[a_{1}(t),
a_{1}^{*}(t)]$), which would also correspond to the physically
unrealistic situation of the $(0,0)$ level being completely filled, as we already discussed in the comments following Eq. (\ref{4f}) in the main text.
On the other hand, we see that the values on the circle of radius
$\ga_{1}(\theta_{0})$ are indeed forbidden for the amplitude
$A_{1}$ since in that case $U^{2}
= - \frac{1}{3} \cos^{2} \theta_{0} < 0$, which is not acceptable
since  $U$ is real.

Consistently, the conservation
law (\ref{15}) and the reality condition for $U$ require that
$|a_{1}(t)|^{2} \leq \frac{1}{3}\sin^{2}\theta_{0}$, which is below $\ga_{1}^{2}(\theta_{0})$. Below such a threshold is also the value
$\frac{1}{6} \sin^{2} \theta_{0}$ taken by $A^{2}_{1}$ when
$|a_{0}(t)|^{2} = \ga_{0}^{2}(\theta_{0})$.
The conclusion is that the field $a_{1}(t)$
is  a massive field with (real) mass
(pulsation) $\si^{2} =2~{\Om}^{2}(1 + \sin^{2} \theta_{0})$.

We consider now the em mode $u(t)$. The potential is
\bea \lab{20}  V[u(t), u^{*}(t)] &=&   3 {\Om}^{2} (|u(t)|^{2} + \frac{1}{3} \cos
2\theta_{0})^{2} \\ \nonumber
&=& \mu^{2}|u(t)|^{2} +
3~{\Om}^{2}|u(t)|^{4} + ~\frac{1}{3} {\Om}^{2} \cos^{2}
2\theta_{0},
\eea
where we put $\mu^{2} = 2~{\Om}^{2}\cos 2\theta_{0}$. For $\theta_{0} \le \frac{\pi}{4}$ ($\mu^{2} \geq 0$), $V[u(t), u^{*}(t)]$ is a paraboloid potential  with
cylindrical  symmetry about an axis orthogonal to the plane
$(u_{R}(t),u_{I}(t))$, with $u_{R}(t)$
and $u_{I}(t)$ denoting the real and the imaginary component,
respectively, of the $u(t)$ field. The minimum (the ground state) is at $u(t) = 0$, for any $t$.
However, we have seen above that in its time evolution the system runs away from
$u(0) = 0$.  Thus $u(t) = 0$ is not acceptable at any $t$ since the system would run away from it at any $t$ and therefore it cannot be a stable ground state. This means that
consistency with the dynamics of the modes $a_{0}(t)$ and $a_{1}(t)$ excludes the possibility $\theta_{0} \leq \frac{\pi}{4}$.

Consistency is recovered
for $\theta_{0} > \frac{\pi}{4}$ ($\mu^{2} < 0$). The potential is no more a paraboloid, but it has a
relative maximum at $u(t)  = 0$, for any $t$, and a (continuum) set of minima (ground states) on the circle of squared radius $|u(t)|^{2}$ in the $(u_{R}(t),u_{I}(t))$
plane:
\be \lab{23}  |u(t)|^{2}  = - \frac{1}{3} \cos 2\theta_{0} = -
\frac{\mu^{2}}{6{\Om}^{2}} \equiv v^{2}(\theta_{0}) ~, ~~~~
 \theta_{0} > \frac{\pi}{4}, \ee
which is a time-independent value.
In a self-consistent way the lower bound $\frac{\pi}{4}$ for $\theta_{0}$ is thus dynamically imposed. Eq.~(\ref{23}) shows that the specific
value of $\theta_{0}$, on which the details of the dynamics depend, is determined by the value of $|u(t)|^{2}$. The fact that $u_{0}(t) =0$, for any $t$, is a relative maximum for the potential is now consistent with the system dynamics evolving
away from it towards the ``limit cycle'' Eq.~(\ref{23}), similarly to the situation for the
$a_{0}(t)$ mode where the system spontaneously evolves away from the
initial conditions.

On the circle of squared radius $|u(t)|^{2}$, the (infinitely many) possible vacua
for the system transform into each other under shifts of
the field $\varphi$: $\varphi \rightarrow \varphi + \al$. One specific ground state
is singled out by fixing the value of the $\varphi$ field.
The ($U(1)$) phase symmetry is broken, and the order parameter
is  given by $v(\theta_{0}) \neq 0$. As usual \cite{ItkZuber}, one can show  that there is
a ``massive'' mode with real mass
$\sqrt{2|\mu^{2}|} = 2\Om \sqrt{|\cos 2\theta_{0}|}$ (a
quasi-periodic mode) and that the field $\varphi(t)$ is
a massless mode playing the role of the NG collective mode.
When Eq.~(\ref{23}) holds, i.e., when $\theta_{0}
> \frac{\pi}{4}$, $A_{1}^{2} = \frac{1}{6} < \frac{1}{3}\sin^{2}\theta_{0}$ (lying below
the upper bound), and $A_{0}^{2} =
\frac{1}{2},$ the requirement that $|a_0(t)|^2 > \cos^{2}\theta_{0}$ is satisfied.

\bibliography{PRE_12Dec_PK.bbl}

\end{document}